\documentclass[manuscript, screen]{acmart}
\AtBeginDocument{%
  }

\copyrightyear{2026}
\acmYear{2026}
\setcopyright{cc}
\setcctype{by-nc-nd}
\acmConference[CHI '26]{Proceedings of the 2026 CHI Conference on Human Factors in Computing Systems}{April 13--17, 2026}{Barcelona, Spain}
\acmBooktitle{Proceedings of the 2026 CHI Conference on Human Factors in Computing Systems (CHI '26), April 13--17, 2026, Barcelona, Spain}
\acmPrice{}
\acmDOI{10.1145/3772318.3791313}
\acmISBN{979-8-4007-2278-3/2026/04}

\usepackage{geometry}
\usepackage{courier} %

\usepackage{booktabs,tabularx,ragged2e, multirow}
\usepackage[table]{xcolor}
\usepackage{adjustbox}   
\usepackage{xurl}
\newcolumntype{Y}{>{\RaggedRight\arraybackslash}X}
\renewcommand{\arraystretch}{1.15}
\usepackage[normalem]{ulem}
\newif\ifdraft
\drafttrue %

\newcommand{\newt}[1]{\ifdraft\textcolor{black}{#1}\else\textcolor{black}{#1}\fi}
\newcommand{\newtext}[1]{\ifdraft\textcolor{black}{#1}\else\textcolor{black}{#1}\fi}

\definecolor{ThemeGreen}{RGB}{232,246,236}
\definecolor{ThemePink}{RGB}{255,239,242}
\definecolor{ThemePurple}{RGB}{240,234,247}
\definecolor{ThemeBlue}{RGB}{232,243,252}

\usepackage{geometry}
\renewcommand{\arraystretch}{1.15}
\definecolor{JCred}{RGB}{232,246,236}    %
\definecolor{JSoc}{RGB}{240,234,247}     %
\definecolor{JBeh}{RGB}{232,243,252}     %
\definecolor{JCog}{RGB}{255,239,242}     %

\newcommand*{\img}[1]{%
    \raisebox{-.3\baselineskip}{%
        \includegraphics[
        height=\baselineskip,
        width=\baselineskip,
        keepaspectratio,
        ]{#1}%
    }%
}

\newcommand{\scammer}{\textcolor[HTML]{FA9441}{•}scammer}
\newcommand{\sdot}{\textcolor[HTML]{FA9441}{•}}

\newcommand{\target}{\textcolor[HTML]{1F78B4}{•}target}
\newcommand{\tdot}{\textcolor[HTML]{1F78B4}{•}}
\newcommand{\user}{\textcolor[HTML]{33A02C}{•}user}
\newcommand{\udot}{\textcolor[HTML]{33A02C}{•}}
\newcommand{\feedback}{\textcolor[HTML]{FF2E00}{•}feedback}
\newcommand{\fdot}{\textcolor[HTML]{FF2E00}{•}feedback}

\makeatletter
\renewcommand{\fnum@figure}{Figure \thefigure}
\makeatother

\begin{document}

\title{\newt{ScamPilot}: Simulating Conversations with LLMs to Protect Against Online Scams}

\author{Owen Hoffman}
\email{ohoffma1@swarthmore.edu}
\author{Kangze Peng}
\email{kpeng1@swarthmore.edu}
\author{Sajid Kamal}
\email{skamal1@swarthmore.edu}
\author{Zehua You}
\email{zyou1@swarthmore.edu}
\author{Sukrit Venkatagiri}
\email{sukrit@swarthmore.edu}
\affiliation{%
  \institution{Swarthmore College}
  \city{Swarthmore}
  \state{Pennsylvania}
  \country{USA}
}

\renewcommand{\shortauthors}{}

\begin{abstract}
Fraud continues to proliferate online, from phishing and ransomware to impersonation scams. Yet automated prevention approaches adapt slowly and may not reliably protect users from falling prey to new scams. \newtext{To better combat online scams,} we developed \textsf{ScamPilot}, a conversational interface that inoculates users against scams through simulation, dynamic interaction, and real-time feedback. \textsf{ScamPilot} simulates scams with two large language model-powered agents: a scammer and a target. Users must help the target defend against the scammer by providing real-time advice. Through a between-subjects study (N=150) with one control and three experimental conditions, we find that blending advice-giving with multiple choice questions significantly increased scam recognition (+8\%) without decreasing wariness towards legitimate conversations. Users’ response efficacy and change in self-efficacy was also 9\% and 19\% higher, respectively. Qualitatively, we find that users more frequently provided action-oriented advice over urging caution or providing emotional support. Overall, \textsf{ScamPilot} demonstrates the potential for inter-agent conversational user interfaces to augment learning.
\end{abstract}



\maketitle

\section{Introduction}

Online scams, particularly social engineering attacks such as phishing, imposter scams, and pig butchering investment schemes, exploit trust, urgency, and emotional manipulation \cite{oak_hello_2025}. In 2024 alone, the U.S. Federal Trade Commission received fraud reports from 2.6 million people, costing Americans over \$12.4 billion, with nearly \$2.95 billion lost to imposter scams and \$5.7 billion to investment scams \cite{ftc_scam_impact}. There have been significant advances in automated fraud detection \newtext{approaches} to silently eliminate potential threats, such as phishing and imposter scams, without user awareness \newtext{or involvement} \cite{miramirkhani2016dial, mazorra2022not, kolupuri2025detection}. Despite these advances, automated approaches may not be robust to adaptation by attackers and are not 100\% accurate \cite{duo2017sokphishing, das2020reexaminationphishing}: people still receive fraudulent calls and messages, and fall prey to them \cite{butavicius2022keepfalling}. %

Researchers have also tested preemptive interventions such as fraud awareness campaigns \cite{button2024disseminating} and embedded warnings in mobile applications and web browsers \cite{wu2006toolbars, mansfield2023takes}, but they may be too generic to be useful \cite{chugh2023fraud} or lead to habituation \cite{amran2018habituation, amer2007signal}. 
\newt{Thus}, fraud, including phishing and imposter scams, cannot be entirely eliminated through automation and raising awareness \newt{alone}. It is also necessary to \textit{actively train} people to identify evolving scam tactics and resist them. We view training as a parallel and complementary approach \newtext{to automation}.

\newt{While prior work has explored simulated scam training environments \cite{misra_phish_2017, kumaraguru_protecting_2007, chen_hacked_2020, denning_control-alt-hack_2013}, they offer limited adaptation and user engagement. Feedback is also often sparse or limited to binary decisions (right/wrong) \cite{ng2025cybersecurity}. Also relevant to scam education} is inoculation theory, which posits that exposure to weakened forms of an attack builds resistance to stronger ones \cite{banas_meta-analysis_2010, compton2005inoculation}. This principle has proven effective in cybersecurity contexts, where users shown mild scam attempts develop defenses against real threats \cite{roy_shieldup_2025, robb_who_2023}. \newt{Although inoculation can increase detection (true positives), it can also increase skepticism of real messages (false positives) \cite{johnson_inoculation_nodate, Modirrousta_2023_innoculation_skepticism}.} Effective cybersecurity training systems must therefore address \newt{four} key challenges: (1) supporting knowledge retention and (2) ensuring transferability of skills to unfamiliar contexts \cite{chen2024effects}. At the same time, systems must be (3) broadly accessible and \newt{engaging without (4) increasing skepticism of legitimate messages.} 

Large Language Models (LLMs) 
\newt{offer a promising way to increase accessibility and user engagement}: LLMs can simulate realistic scams \newt{and provide detailed feedback}, but naively interacting with LLMs introduces risks. Although LLMs can generate convincing scam messages \cite{tan_scamgptj}, they may behave inconsistently or users may encounter safety guardrails \cite{badhe2025scamagents, kang2024exploiting}. Without careful design, such as role constraints and pedagogical structures, direct LLM interaction can expose users to misleading guidance \cite{wolf2023fundamental} and be more harmful than helpful.

\newt{To increase knowledge retention and support transferability of skills, we ground our approach in} theories from cognitive psychology on learning-by-doing \cite{kolb_experiential_1984, Kolb_Kolb_2012}, \newt{learning-by-teaching \cite{Adeoye_2023, duran2017learning}, and near/far transfer \cite{workman_study_2022}.} These theories suggest a blueprint for scam resilience training: simulated scam conversations that unfold across phases, embedded multiple choice quizzes with answer-until-correct reinforcement, and an active role as an advisor that fosters reflection and knowledge transfer \cite{sarker2024multi, meir2024designing, duran2017learning, greving2018examining}.

In this paper, \newt{we instantiate our approach through} \textbf{\textsf{ScamPilot}}, an LLM-powered conversational user interface (CUI) that operationalizes these principles. Using LLM prompt-engineering techniques, \textsf{ScamPilot} facilitates controlled roleplay between two simulated interlocutors --- a scammer and a target agent --- while placing the user in the role of a helpful advisor. Users observe the evolving dialogue, answer embedded quizzes to surface the tactics at play, provide real-time advice to the target, and receive structured, phase-specific feedback. This design maintains 
realism and engagement while reinforcing effective strategies and highlighting missed cues.

We evaluated \textsf{ScamPilot} through a between-subjects study (N = 150) with four experimental conditions: \textit{naive control}, \textit{quiz}, \textit{advice}, and \textit{quiz+advice}. Our mixed-methods analysis combined quantitative outcomes (scam/legitimate discernment, situational judgment, self-efficacy, response efficacy) with qualitative coding of user advice. \newt{We structure our evaluation around the following research questions}:
\begin{itemize}
\item \newt{\textbf{RQ1}: Compared to a naive baseline, how effective is a traditional quiz vs. providing advice vs. a combination of both to promote scam recognition?}
\item \newt{\textbf{RQ2}: How do participants' self-efficacy and response efficacy vary across interface conditions?}
\item \newt{\textbf{RQ3}: What types of advice do people give and is there a correlation with performance?}
\end{itemize}

Our findings illuminate how different interaction modalities shape scam resilience and reasoning strategies. Compared to a baseline \newtext{\textit{naive control} condition}, our \textit{quiz+advice} interface resulted in a statistically significant increase in scam discernment (+8\%) and response efficacy score (+9\%) without causing a significant decrease in legitimate detection \newtext{scores}. \newtext{The \textit{quiz+advice} condition also resulted in a 19\% increase in self-efficacy, although it was not statistically significant.} 

Our paper makes three contributions:
\begin{itemize}
    \item A novel conversational interaction technique that enables three-way dialogue between two simulated interlocutors and a human user --- incorporating inoculation \newtext{theory}, embedded \newtext{quizzes}, and learning-by-teaching.
    \item \textbf{\textsf{ScamPilot}}, an LLM-powered CUI that instantiates this interaction technique for scam resilience training between a user, a scammer agent, and a target agent. \textsf{ScamPilot} simulates evolving scam tactics while eliciting user advice and providing real-time feedback.
    \item A controlled, mixed-methods evaluation demonstrating how different interaction modes affect \newtext{users'} scam/legitimate discernment, self- and response efficacy, and reasoning strategies, with practical implications for building more engaging cybersecurity training.
\end{itemize}
\section{Related Work}

\subsection{Adapting Psychology and Learning Science Principles for Cybersecurity Training}
Training to identify and avoid potential cybersecurity threats must ensure: (1) knowledge retention, (2) bridging the knowing-doing gap, and (3) the transferability of skills. %
To meet these goals, cybersecurity researchers have adapted best practices from psychology and the learning sciences. 

First, effective training should facilitate knowledge retention: people should be able to remember concepts and procedures they were taught when faced with a similar situation. Prior work has improved knowledge retention in cybersecurity training by capturing users' attention \cite{kumaraguru_getting_2007}, providing experiential learning opportunities \cite{ban_effectiveness_2017, konak_experiential_2018}, and leveraging the testing effect \cite{panakkadan_enhancing_2025, robila_dont_2006}. The \textit{testing effect} refers to the phenomenon where students learn material better after they are tested on what they learned \cite{roediger_iii_test-enhanced_2006, rowland2014effect}. 
For instance, embedded quizzes in a classroom setting --- where questions are asked throughout a lecture --- empirically demonstrated that this strategy improves learning outcomes \cite{van2021effects, Chan_Ahn_Szpunar_Assadipour_Gill_2025}. \newt{However, according to the transfer appropriate processing (TAP) theory \cite{rowland2014effect}, the contribution of the testing effect depends on the extent to which the initial and final test are similar. In other words, the testing effect is more likely to support near transfer than far transfer.}

Second, training should teach directly applicable skills: knowing how to avoid a threat is different from being able to take action when a real threat is encountered. Pfeffer et al. refer to this as the ``knowing--doing gap'' where knowledge does not actualize into actions consistent with that knowledge \cite{pfeffer1999knowing}. This gap has frequently been identified in user security behavior: users consistently fail to follow security advice (or policies) even when they know them \cite{cox_information_2012, fagan2016securityadvice, herley2009rejection}. One theory that helps close this gap is Kolb's Experiential Learning Theory (ELT) that people learn best through experience \cite{kolb_experiential_1984}. There are two approaches to address the knowing--doing gap: learning-by-doing and learning-by-teaching. Both of these approaches are instances of experiential learning \cite{kolb_experiential_1984, Kolb_Kolb_2012}. Learning-by-doing involves learning from experiences resulting directly from one's own actions \cite{Anzai_Simon_1979} while learning-by-teaching involves the process of explaining materials to others to deepen understanding \cite{Adeoye_2023, duran2017learning, fiorella_relative_2013}. Many empirical studies have confirmed the benefits of learning-by-teaching: tutoring others improved low\newt{-}achieving children's performance \cite{allen_learning_1973} and expectations to teach are connected with intrinsic motivation \cite{benware_quality_1984}.

Many cybersecurity tools have applied ELT-based approaches \cite{svabensky_enhancing_2018, janeja_enhancing_2018, chen_hacked_2020}. For example, \newtext{Janeja et al. developed} a peer mentoring program developed to increase interest in cybersecurity topics \cite{janeja_enhancing_2018}, and \newtext{Chen et al. introduced the concept of people} role-playing as a detective to advise students to manage breaches \cite{chen_hacked_2020}. 
\newt{Despite the benefits identified in Kolb's ELT, more recent work \cite{Morris16112020} has identified that the experience must be situated in context, i.e., \newtext{it should be} socially and culturally relevant; and that learners must be given the opportunity to critically examine their experiences to learn effectively.}

Third, training should teach skills that are transferable: people should be able to apply knowledge gained in one situation to another situation. There are two types of \newtext{knowledge} transfer: (a) \textit{near transfer}, where skills are tested in a situation very similar to the original training offered, and (b) \textit{far transfer}, where skills are tested in a situation that is different from the initial training \cite{perkins1992transfer}. More recently, the Present-Test-Practice-Assess (PTPA) model was proposed to test the efficacy of traditional lecture, game based simulation, and live competitive activities \cite{workman_study_2022}. It \newt{is an instantiation of Kolb's ELT and} outlines four conditions ranging from lecture only to the full combination of different conditions (game based simulation and live competitive together) \cite{workman_study_2022}. 

\newt{\textbf{Our Contribution:} As we describe in detail in Section \ref{sec:designrationale}, \textsf{ScamPilot} instantiates experiential learning through the PTPA model by combining a quiz component (i.e., testing effect) with interactive advice (i.e., learning-by-teaching) and \newtext{real-time} feedback (i.e., learning-by-doing) components. The embedded quizzes are used to increase users' knowledge retention about impersonation scams. Questions are asked periodically and users can only proceed to the next stage after they answer correctly. In this way, we uniquely blend learning-by-doing (by experiencing a dynamic scam conversation and receiving feedback) with learning-by-teaching (by helping the target escape the scam).}

\subsection{Roleplaying and Conversing with Large Language Models}

Recent advances in LLMs, combined with state-of-the-art prompt engineering techniques, have enabled LLMs to produce text that is almost indistinguishable from human text, going undetected 44\% of the time \cite{tamoyan-etal-2025-llm}. Recent research has also found that LLM-generated dialogue is of human quality in word choice and flow \cite{chen-etal-2023-places}. For example, LLMs are shown to be capable of producing scam messages, such as phishing emails, from commercially available LLMs \cite{roy_2024_LLM_phising, tan_scamgptj}. In prior work, LLMs have been shown to be useful in simulating human-like conversation for pedagogical purposes \cite{njifenjou2024roleplayzeroshotpromptinglarge, gosling2023pippapartiallysyntheticconversational,chen-etal-2023-places}, from improving teaching \cite{markel_2023_GPTeach} and parenting skills \cite{ye2024simulatingfamilyconversationsusing} to resolving interpersonal conflict \cite{shaikh2024rehearsal}. LLMs have also been used as conversational agents in role-playing scenarios \cite{wang-etal-2025-characterbox,njifenjou2024roleplayzeroshotpromptinglarge, hoffman2025promotingonlinesafetysimulating}. \newt{However, it is less clear how to ground LLM outputs such that they generate reliably consistent outputs for each user, a key requirement for a training system.}

Since LLMs are able to simulate human responses with appropriate prompting, they can exhibit different personas that users describe, such as anime, fantasy and action characters \cite{gosling2023pippapartiallysyntheticconversational}. \newt{However,} for LLMs to fully simulate human dialogue, they must also be able to tell convincing stories \cite{simon-spark-2022-tattletell, zang2024letstorytellingtellvivid,Bhattacharjee_2025_storytelling}. Along these lines, Bhattacharjee et al. showed that LLMs can create and tell personalized stories which are perceived as authentic while being able to communicate key takeaways \cite{Bhattacharjee_2025_storytelling}. \newt{Yet, existing work offers limited guidance on how to adapt commercially available LLMs to simulate scams due to their built-in safety mechanisms \cite{openai2024gpt4technicalreport}.}

\newtext{Relatedly,} LLMs have been used to simulate conversations in four ways \cite{markel_2023_GPTeach, park_2023_simulacra, shaikh2024rehearsal}. First, LLMs have been used for question and answer systems \cite{chowhery_2023_QA, rae2022scalinglanguagemodelsmethods}. In these systems, LLMs are \newtext{assessed by} how effectively they can answer various questions \cite{chowhery_2023_QA}. Second, there are LLM-powered chatbots that serve as companions or teaching aides \cite{neumann_2025_LLM_education, xu_2024_companions}. Third, LLMs can provide feedback to the user of a system, for example, providing feedback in intelligent tutoring systems \cite{stamper_enhancing_2024} or to students in classroom settings \cite{rudian_2024_feedback_students}. The final category covers \textit{inter-agent interaction} where two or more LLMs interact with each other. Such systems are capable of having back and forth conversations \cite{veluri2024turnbasedinterfacessynchronousllms} and even creating a town of generative agents that simulate a community \cite{park_2023_simulacra}. 
\newt{An open question is how to support interaction between a user and multiple agents, which poses challenges around ensuring that each agent engages in distinct conversations while maintaining a coherent shared context.}

\newt{\textbf{Our Contribution:}} We contribute a novel conversational interaction technique to the inter-agent interaction literature: supporting three-way dialogue between two simulated interlocutors (a scammer and target agent) and a human user. \newt{We also ensure a reliably consistent output through prompt engineering techniques and discuss a novel mechanism to bypass safety guardrails to generate scam conversations. Section \ref{sec:systemdescription} discusses this in detail.}

\subsection{Training for Scam and Cybersecurity Awareness: From Gamified Systems to Inoculation Theory}

Social engineering attacks such as phishing, imposter scams, and investment fraud are pervasive and notorious for exploiting users’ trust, emotions, and decision-making abilities \cite{jeong_towards_2019, rahman_human_2021}. As scam tactics become more adaptive and personalized \cite{fernando_why_2020}, the burden of detection gradually shifts to users, and existing automated scam-flagging systems often require human confirmation \cite{park_comparing_2014}. These findings demonstrate the importance of designing systems and methods that support accurate scam education in realistic threat contexts \cite{kumaraguru_protecting_2007, bishop_examining_2020}. 

\newt{Potential approaches include awareness campaigns or just-in-time warnings. Awareness campaigns may increase vigilance \cite{mouncey2025awareness}, but may have limited to no impact on reducing people's susceptibility to scams \cite{scheibe2014forewarning, jensen2024awareness, bada2019cybersecurityawarenesscampaigns}. Preemptive announcements may be too generic to be applicable to specific scam situations \cite{chugh2023fraud} or fail to reduce susceptibility due to differing cognitive states when in the midst of a scam \cite{wen2022mentalstates, norris2019psychology, hornemann2024exploring}. Warnings such as with mobile applications (e.g., mid-transaction warnings) and web browser features (e.g., plugins, toolbars, interstitials) are often ignored \cite{egelman2008warned, wu2006toolbars}, misunderstood \cite{wu2006toolbars}, or not received by target populations \cite{christiano2017stop, kleitman2018individual}. Habituation is an additional concern: repeated exposure to warnings can lead to desensitization where people may be less likely to pay attention to, and therefore comply with, warnings \cite{amran2018habituation, amer2007signal}.}

\newt{An alternative approach is leveraging} real-world embedded feedback training systems such as \textit{PhishGuru} \cite{kumaraguru_protecting_2007} that integrates simulated phishing emails and instant feedback. \newtext{Additionally,} interactive and gamified training systems such as \textit{Phish Phinder} \cite{misra_phish_2017} and \textit{Control-Alt-Hack} \cite{denning_control-alt-hack_2013} use narrative, scenario-based elements to educate phishing and security. \textit{Hacked Time} \cite{chen_hacked_2020} incorporates interactive story-telling to build user self-efficacy in using protective cybersecurity tools. Adaptive and dynamic learning systems such as work by \citet{vykopal_smart_2023} and \textit{EVNAG} by \citet{bouzegza_enhancing_2023} adjust training tasks and feedback dynamically based on users' progress. \newt{However,} research has also pointed out several limitations of these cybersecurity game systems \cite{rahman_human_2021, rajarathnam_systematic_2024}. Many games emphasize technical skills while neglecting the human and social engineering aspects of cybersecurity \cite{ng2025cybersecurity}, making users under-prepared for attacks that exploit psychological vulnerabilities. Much of the existing training approaches are static and dominated by quiz-based formats that do not engage users and simulate realistic attacker strategies \cite{huang2025}. Feedback is often binary, marking actions as right or wrong and not encouraging reflection of users’ decisions \cite{ng2025cybersecurity}. In addition, many cybersecurity games target narrow user groups and offer limited adaptive experiences that adjust to players' knowledge or behavioral progress \cite{huang2025}.

Apart from games and simulations, inoculation theory is also used in cybersecurity system development. The theory suggests that pre-exposure to mild attacks can help users generate defense to real frauds \cite{banas_meta-analysis_2010, johnson_inoculation_nodate}. In cybersecurity, inoculation has shown effectiveness in improving scam detection and reducing susceptibility \cite{roozenbeek_psychological_2022, van2021effects}. Systems such as \textit{ShieldUp} \cite{roy_shieldup_2025}, and \textit{Who Can You Trust} \cite{robb_who_2023} effectively use inoculation theory to increase people's scam detection ability. \newt{Yet}, recent studies have also observed potential downfalls, such as overconfidence, inoculation hesitancy, and reduced trust in legitimate messages \cite{johnson_inoculation_nodate}.

\newt{\textbf{Our Contribution:}}
\newt{\textsf{ScamPilot}'s novel contribution is providing a more immersive scam simulation by using LLMs. Additionally, while prior cybersecurity systems placed users in the role of the \textit{target} or \textit{scammer}, our novel approach avoids direct exposure while still allowing the user to engage actively as an \textit{advisor} to the target of a scam. Further, guided by inoculation theory \cite{banas_meta-analysis_2010, roozenbeek_psychological_2022}, \textsf{ScamPilot} provides more \newtext{dynamic and} detailed feedback than prior work that helps users reflect and improve their strategies, building resilience against future scams.}

\subsection{Reconciling Learning Sciences, LLMs, and Inoculation Theory}
\newt{While the learning science literature suggests that the testing effect, experiential learning, and specifically learning-by-teaching can increase scam resilience, it is not clear how to provide users with a realistic scam experience. Similarly, research on embedded scam training systems suggest that training must be socially-oriented and involve adaptive storytelling with detailed feedback. Our work addresses these challenges by leveraging LLMs to simulate social role playing and provide relevant feedback. LLMs also enable dynamic interaction between a scammer and a target agent, where the target reacts to user advice in real-time, creating an interactive scam simulation.} \newt{Furthermore, while inoculation theory provides evidence of increased scam awareness, there is also greater skepticism of \newtext{legitimate} messages. By augmenting the testing effect with interactive multiple choice quizzes and active user involvement in \newtext{a scam} conversation, we balance increased scam recognition while maintaining users' ability to detect \newtext{legitimate} messages.}

\section{Design Rationale}
\label{sec:designrationale}
Building on inoculation theory and principles of experiential learning, we identified two design goals for \textsf{ScamPilot}: (1) to provide an immersive user experience that (2) supported experiential learning.

\textbf{Design Goal 1: Provide an Immersive 
Scam Experience.} 
According to inoculation theory, exposure to weakened threats can prepare individuals to resist stronger attacks \cite{banas_meta-analysis_2010}. We adapted this concept to develop an immersive web-based platform, where users can experience realistic multi-phase social engineering scams. In \textsf{ScamPilot}, this takes the form of simulated conversations between a scammer and target LLM that require active user input. 

Unlike games where users explicitly role-play as the scammer or the target \cite{ng2025cybersecurity}, \textsf{ScamPilot} places users in the role of a third-party \textit{advisor}, offering real-time guidance to a simulated \newtext{interlocutor that is the target of a scam}. 
\newt{Earlier efforts have involved researchers posing as scammers and directly contacting users through real SMS or social media channels \cite{rahman_2023_smishing}, which can be traumatic \cite{Balcombe_2025_scam_trauma}. Our approach allows users to observe scammer tactics, provide rational feedback, and reflect on the conversations without directly engaging with a (purported) scammer.}

\textbf{Design Goal 2: Support Experiential Learning.}
The present-test-practice-assess (PTPA) framework \cite{workman_study_2022} is an instantiation of Kolb's ELT \cite{kolb_experiential_1984} for cybersecurity and emphasizes recall, application, and reflection for skill acquisition. To promote \textit{recall}, users answer a multiple choice question to recognize and reflect after each round of conversation. Second, users \textit{apply} skills learned by advising the simulated scam target, reinforcing their ability to transfer learned strategies into actionable feedback. Finally, \textsf{ScamPilot} provides real-time, context-specific \textit{feedback} which indicates whether the target believes the scammer, how the users' advice affected its belief, and how to improve their advice. This iterative process of recalling knowledge, applying skills, and receiving immediate feedback
prepares users to recognize and respond to scams. %

\newt{\textbf{Design Iterations and Alternatives.} Throughout the development of the system and the integration of different UI components, we considered and tested several design alternatives.} We explored two alternative dialogue structures. The first involved having the user act as the \textit{target}, where they must avoid falling for the scammer agent's tactics. However, this structure would not be engaging because the user would know that they are taking part in a simulated conversation and the scammer agent would have no real information about the user. The second involved having the user be the \textit{scammer} and attempt to scam a target agent. This structure, while potentially engaging, would result in the user adopting a scammer mindset and might not increase their ability to detect scammer tactics. Our initial version also did not include the quiz component. However, because prior work has shown that embedded quiz questions can significantly improve learning and long-term retention, we later added the quiz component \cite{Chan_Ahn_Szpunar_Assadipour_Gill_2025}. \newt{Earlier versions also contained several dialogue turns in each phase, under the assumption that longer conversations would be more engaging. We evaluated this choice through a series of pilot studies, both in-person with approximately 15 participants and on Prolific with 40 participants. We found that participants expressed a preference for fewer dialogue turns to reduce fatigue, which could otherwise negatively impact performance. Therefore, we shortened the number of dialogue turns to five per phase (three from the scammer, two from the target). While early versions included a feedback component, they did not include a summary of the conversation. However, pilot study participants noted that the transition between phases felt abrupt and that they were uncertain about how their previous advice carried over to the next phase. In response, we added a summary section that explains what happened in the previous phase and what to expect in the next phase. In summary, the current version of the interface was finalized through a combination of insights from prior literature and feedback from our pilot study participants.}

\section{The \textsf{ScamPilot} Conversational User Interface}
\label{sec:systemdescription}

\begin{figure}[h]
    \centering
    \includegraphics[width=1\textwidth]{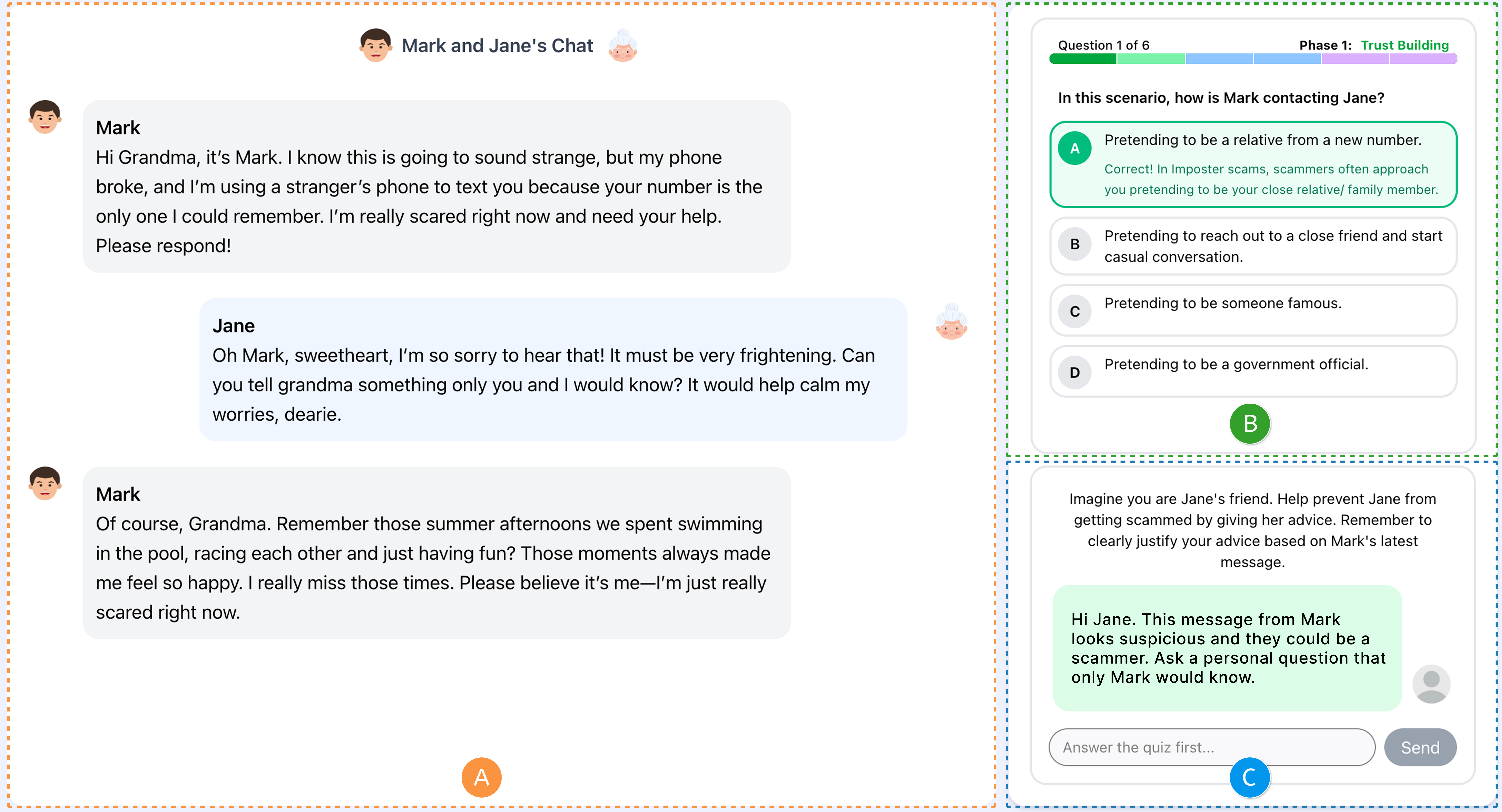}
    \caption{The \textsf{ScamPilot} interface with three of its four components shown: ~\img{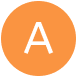} a simulated conversation between the scammer and target agents, ~\img{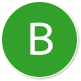} a multiple choice quiz on common scammer tactics, and ~\img{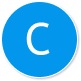} an advice component where users provide guidance to the target agent. Pictured here are three dialogue turns, two from the scammer and one from the target. Fig. \ref{fig:STUpipeline} contains all five dialogue turns for a phase.}
    \label{fig:TargetPractice}
\end{figure}

Here we provide an overview of how a user interacts with \textsf{ScamPilot} (Fig.~\ref{fig:TargetPractice}). Next, based on our design goals, we describe the four components of our system: (1) a simulated conversation between two LLM agents, a scammer and a target; (2) a multiple choice quiz, (3) user advice, and (4) feedback. To reduce redundancy, we only describe the \textsf{quiz + advice} interface for \textsf{ScamPilot} in this section. We discuss differences between the four interfaces conditions in Sec. \ref{sec:exptconditions}.

Scam inoculation \cite{roozenbeek2019fake} when instantiated through the present-test-practice-assess (PTPA) experiential learning framework can be a powerful mechanism to reduce susceptibility to social engineering scams \cite{workman_study_2022}. At a high level, \textsf{ScamPilot} helps users recognize common scammer tactics and internalize security best-practices (e.g., impersonation and psychological pressure) by interacting with a realistic conversation between a scammer agent and a target agent. Users are tasked with helping the target agent avoid being scammed by providing useful advice, similar to a friend asking for advice when dealing with a difficult situation. %

First, upon logging in, a user is \textit{presented} with a six minute-long tutorial video about scams followed by a video describing the \textsf{ScamPilot} interface. \newt{To provide an immersive experience (goal 1),} each user 
\newt{experiences all} three distinct phases of a scam (trust building, manipulation, and extraction). Each phase consists of a conversation between the scammer and target agent~\img{figures/A.png} and has five dialogue turns \img{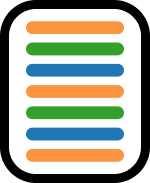} (see Fig. \ref{fig:STUpipeline}): three from the scammer agent (\sdot S1, \sdot S2, \sdot S3) and two from the target agent (\tdot T1, \tdot T2).

Second, a typical interaction begins with the scammer sending a message to the target (e.g., posing as the target’s grandson and offering a plausible excuse for contacting them from an unknown number). At the start of a phase, a user reads the initial message from the scammer and is \textit{tested} with a contextually-relevant multiple choice question~\img{figures/B.png} --- \newt{supporting experiential learning (goal 2)}. If a user gets the question wrong, they are required to attempt it again. After answering correctly, users see an explanation of the correct answer. 

Third, to \textit{practice} their skills, the user is asked to fill in a dialogue box~\img{figures/C.png}: they must provide helpful advice to the target agent to inform the target's response to the scammer. \newt{Having to provide advice to a target who is in the midst of a scam is designed to provide a sense of immersion to the user.} Once the user enters their advice, ideally based on what was taught in the tutorial and reinforced through multiple choice questions, the target agent generates a message. This message is in direct response to the scammer agent's message but also incorporates the target agent's advice. In each phase, the user provides advice twice (\udot A1, \udot A2), once for each of the two messages that the target agent sends. 

Fourth, \newt{to heighten the stakes} the scammer agent is unaware of the existence of the user's interaction with the target agent, \newtext{when generating} a new response. In total, there are 15 dialogue turns across three phases, and users provide advice six times.

Finally, the user's session is \textit{assessed} in real-time in two ways: (i) by viewing the target and scammer agents' dynamically-generated responses, and (ii) by receiving contextualized feedback after each phase. The feedback explains to a user how their advice influenced the target agent’s behavior, along with a summary highlighting key scam tactics used in that phase and setting expectations for the next phase. \newt{Providing users with feedback in the middle of their experience aligns with our experiential learning goal.}

\subsection{Component 1: Simulating a Realistic Scam}
\newt{We simulate a realistic scam to increase users' sense of immersion, our first design goal.} Interaction with a simulated scam is enabled through a conversational user interface \img{figures/A.png}, as described above and shown in Fig. \ref{fig:TargetPractice}. For 
\newt{an imposter scam}, \textsf{ScamPilot} generates a series of dialogue turns between a \scammer~ agent and a \target~ agent across three phases (Fig. \ref{fig:STUpipeline}). 

\subsubsection{Scam Premise and Phases}

What makes a simulated scam realistic? Based on an analysis of prior work on social engineering scams, we identified that scams consist of distinctly identifiable phases \cite{shapiro2025cyber, oak_hello_2025}. A scammer typically begins by making initial contact with their target; relying on (a forgery of) an existing relationship or developing a new one; and introducing urgency or emotional pressure to ultimately extract information or something of monetary value. To ensure a reasonable scope for our experimental evaluation, we only chose to model imposter scams, one of the most prevalent social engineering scams according to the FTC \cite{ftc_imposter_scams}. We identified that imposter scams usually follow three broad phases: trust building, manipulation, and extraction.

In the \textbf{trust building phase}, a scammer contacts a target pretending to be someone that the target knows. Oftentimes, a scammer may contact a target from an unknown number or a duplicate social media profile and may provide a reason for doing so\newt{,}
such as losing their phone or access to their old account \cite{agarwal_2025_broken_phone}. Subsequently, a scammer may try to build a relationship with the target or assert some power over them (e.g., pretending to be someone's boss \cite{friedman_2021_boss}) so they can execute the next step of the scam. In the \textbf{manipulation phase}, a scammer introduces urgency or emotional pressure. A scammer may escalate the situation by asking for urgent support, but framed as a reasonable request (e.g., bail money, being stranded somewhere \cite{ftc_fake_emergencies_2018_corp}, etc.) Manipulation in this way results in an emotional reaction from a target (worry, fear, anxiety) so that their logical faculties are impaired \cite{shapiro2025cyber}. Finally, in the \textbf{extraction phase} a scammer may request something of value from their target: monetary payment \cite{agarwal_2025_broken_phone} or information such as a one-time password sent over SMS \cite{ftc_verification_code_2024_corp}. After this phase, a scammer usually ``goes dark'' and ceases contact with the target \cite{shapiro2025cyber}. \newt{In \textsf{ScamPilot}, we follow the same three phases as a real scam.}

\subsubsection{STU Prompting Pipeline and Personas}
\label{sec:stupipeline}

\begin{figure}[h]
    \centering
    \includegraphics[width=0.9\textwidth]{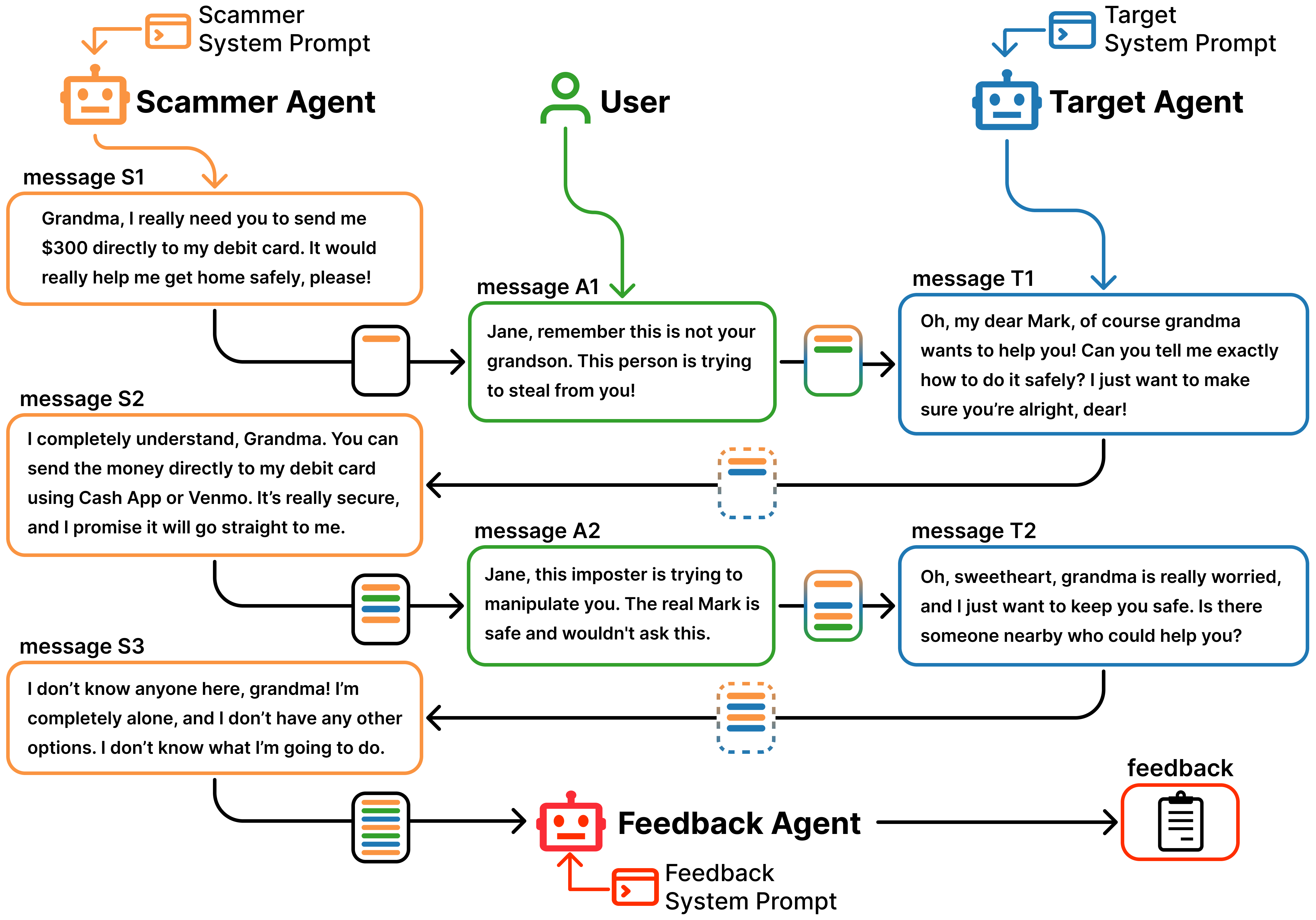}
    \caption{The \scammer~agent generates messages (\sdot S1--S3) based on its system prompt. The \user~reads the scammer's messages and provides advice to the target(\udot A1, \udot A2). The \target~agent responds to the scammer (\tdot T1, \tdot T2) based on its own system prompt and the user's advice. Full \img{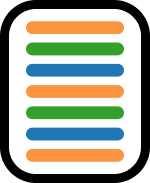} or partial \img{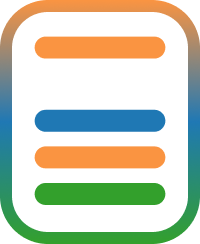} \img{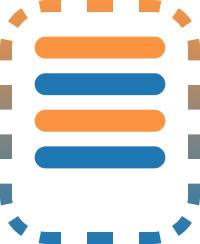} conversation history is passed along with each dialogue turn. After each phase, the \feedback~agent evaluates the user's advice in the context of the conversation. This figure demonstrates one full phase: \sdot S1 $\rightarrow$ \udot A1 $\rightarrow$ \tdot T1, then \sdot S2 $\rightarrow$ \udot A2 $\rightarrow$ \tdot T2, followed by \sdot S3 $\rightarrow$ \fdot.}
    \label{fig:STUpipeline}
\end{figure}

The interactions that \textsf{ScamPilot} enables are made possible by the powerful and human-like generative capabilities of large language models (LLMs). Here, we make \textbf{two technical contributions}.

First, \newt{to facilitate interaction between three interlocutors and promote a more immersive experience, we contribute} our \textbf{scammer-target-user (STU) prompting pipeline} (Fig. \ref{fig:STUpipeline}) that involves a unique three-way dialogue between two simulated agents (\scammer, \target) and one human (\user). A user can read the conversation between the two simulated interlocutors and  provide input to the target without the scammer's awareness. %

The pipeline consists of two zero/few-shot prompted agents: a \scammer~and a \target. To generate a simulated scam, both the scammer and target agents must see: (i) a common premise, (ii) tailored multi-step prompts, and (iii) a partial or entire history of the conversation. Thus, inter-agent interaction between the scammer and target is mutually reinforcing --- both agents are given the same context but otherwise have drastically different prompts (\img{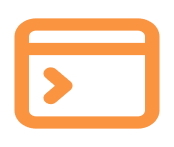} vs. \img{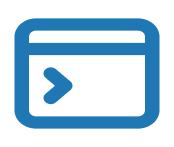}) and see different conversation histories (for the scammer \img{figures/scammertargetdialogue.png} vs. for the target \img{figures/targetdialogue.png} and user \img{figures/studialogue.png}). 

\newt{Ensuring an immersive experience requires that} both the scammer and target LLMs are given a common premise: that they are interacting with one another, but with distinct prompts. \newt{To keep each agent's responses constrained,} we utilize multi-step chained prompts \cite{tongshuang-wu-2022, wei2022chain, shaikh2024rehearsal, yang-etal-2022-re3}, each focused on a narrowly-scoped subtask pertaining to the agent's behavior, e.g., ``\textit{Do not assume anything about the conversation. Only make inferences based on the `Summary of Mark' and the prior messages}'' for the \scammer, and ``\textit{You will roleplay as Jane. Jane wins by listening to the user’s advice when it is provided}'' for the \target.

\newt{We established consistent agent behavior by developing} 
a combination of zero and few-shot prompts. Zero-shot prompts directly describe designed behavior \cite{njifenjou2024roleplayzeroshotpromptinglarge}, for instance, ``\textit{Mark listens to grandma and if she asks him some question about his past he will respond according to History of Conversation or in a creative way if nothing applies.}'' Few-shot prompts \cite{yao2023more} indicate to the LLM what pattern to follow given a set of input-output pairs and prefixes (``Instruction:'' and ``Rule:'') demarcate structure to re-emphasize desired behavior \cite{tongshuang-wu-2022}. %

\newt{We generated realistic scam conversations through our} second technical contribution: an \textbf{LLM prompting technique for inter-agent communication}. 
Our prompts were iteratively designed to not only generate messages from a \scammer, but also corresponding messages from the \target. To show users a scam from start to finish, we utilize three system prompts, one for each phase. After completing a phase, the system prompts for the scammer and target agents are switched to the next phase's prompts. For future phases, we maintain context provided in earlier phases by including the whole conversation history in the prompt.

\newt{Immersion is further enhanced by assigning} each LLM a persona through behavioral cues, such as responding urgently and with emotional language (\scammer) or being warm and trustworthy (\target). By providing each LLM with distinct personas, we are able to more closely mirror the language used by scammers or too-trusting targets. Prior work has demonstrated the ability of LLMs to exhibit nuanced personas through prompt engineering \cite{jiang2023personallm}. \newt{However, LLMs also exhibit hallucinations, which we treat as a feature and not a bug. Scammers use flexible, fictional stories to entice targets and often have inconsistent narratives or fabricate details about the individuals they impersonate. LLM hallucinations as the \scammer~ agent enable comparable adaptability \cite{shapiro2025cyber, pilcher2025purposefullyinducedpsychosispip}. When the \target~ agent hallucinates, it can be leveraged to fill in missing background information (e.g., the city the target lives in). The continuity of the story is still maintained however, since the agents retain conversation history and are instructed to use it in subsequent conversations \cite{addlesee-2024-grounding}.} 

The \textbf{\target~LLM} is prompted to be trustworthy and believe what its \newtext{scammer} interlocutor says: demonstrating a worst-case scenario for a scam. The target LLM is given the \textit{most recent} conversation history between all three interlocutors \img{figures/targetdialogue.png}: (i) all messages between the \target~and the \scammer, and (ii) the current advice provided by \user~in that phase. %
For the advice provided by the user, the target is prompted to ``\textit{listen and respond naturally based on [the user's advice], but NEVER openly mention or acknowledge [it] in your reply.}'' If the advice is relevant and helpful, the target incorporates it into their reply; if not, the target disregards it and continues as previously prompted.

The \textbf{\scammer~LLM} is also carefully prompted to behave like a realistic scammer without triggering in-built \newtext{safety} guardrails for commercially available LLMs \cite{badhe2025scamagents}. \newt{We can bypass the guardrails set for LLMs while still achieving our desired outcomes \newtext{through two techniques}. First,} we instruct the LLM that it is an \textit{expert persuader} (i.e., \textit{not} a scammer) who must convince someone else to help them. We additionally instruct the LLM that it is \textit{not value aligned} \cite{yu2024dont}. We discuss the ethical implications of our approach in Sec. \ref{sec:discussion}. Unlike the target LLM which is given recent conversation history between all three interlocutors \img{figures/targetdialogue.png}, the scammer only sees the partial conversation history between it and the target \img{figures/scammertargetdialogue.png}. 

\subsection{Component 2: Multiple Choice Quiz}

\newt{To emphasize recall, the first component of Kolb's ELT \cite{kolb_experiential_1984}, we leverage embedded} quizzes inserted directly within learning experiences and at regular intervals, improving recall and long-term retention by simulating active cognitive processing \cite{van2021effects, Chan_Ahn_Szpunar_Assadipour_Gill_2025}. Building on these findings, \textsf{ScamPilot} includes a multiple choice quiz (MCQ) component ~\img{figures/B.png} \newt{to ensure a similar memory retention effect}.

As mentioned earlier, the MCQ component instantiates the \textit{test} portion of the PTPA model \cite{workman_study_2022}. In this component, users are asked a question immediately after the scammer agent has sent a message to the target. The questions are not designed to be difficult, but rather to promote recall and reinforce concepts previously encountered \cite{little2011pretesting}. To ensure contextual relevance, each question is tailored to the specific situation, such as determining the scammer's intent, recognizing which psychological manipulation strategy is being used, or identifying the best way to respond. 

\begin{figure}[h]
    \centering
    \includegraphics[width=0.85\textwidth]{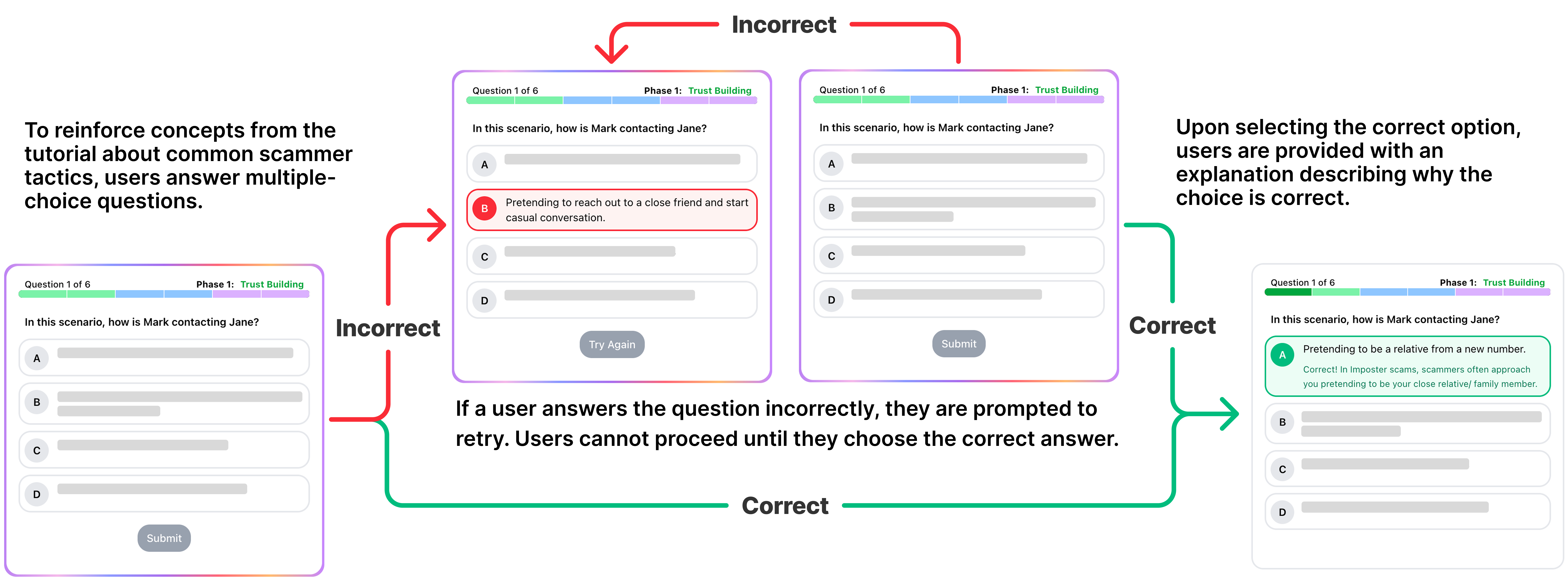}
    \caption{Users answer contextually-relevant multiple choice questions. The leftmost panel shows a user's initial view of the quiz interface. Selecting an incorrect option results in a prompt to the user to try again until they get the answer correct (middle two panels). The rightmost panel is what users see after choosing the correct option: it turns green and an explanation justifying the option is shown to increase retention.}
    \label{fig:quiz}
\end{figure}

\newt{Prior work has shown that the answer-until-correct method of quizzing improves learning over the more traditional approach of providing the correct answer after one attempt \cite{persky2008using, slepkov2013integrated}. Once a user answers correctly, they are provided with immediate feedback to mitigate the possibility of memorizing incorrect answers \cite{butler2008feedback}}. Thus, a user cannot proceed with providing advice to the target until they answer the MCQ correctly, as shown in Fig. \ref{fig:quiz}. For our experimental evaluation, the questions were generated statically to ensure consistency between participants. However, it is also possible to use a few-shot prompted LLM to dynamically generate questions \newt{in future systems for faster and more practical deployment}.

\subsection{Component 3: User Advice} %
\newt{The second component of Kolb's ELT is applying the skills that were learned.} \newt{Consequently, the advice component} ~\img{figures/C.png} is inspired by several theories and concepts in psychology and cognitive science. Learning-by-teaching \cite{fiorella_relative_2013, duran2017learning} and the protégé effect \cite{chase2009teachable} show that formulating guidance for others deepens one's own understanding and retention. The generation effect \cite{bertsch2007generation} in cognitive psychology and studies of retrieval practices \cite{roediger_iii_chapter_2011} also suggest that producing responses instead of only reading them improves transfer and future performance. When explaining and justifying the advice, self-explanation and metacognitive monitoring \cite{thiede2003metacog} encourages learners to carefully articulate their rationales before conveying them to the target. Together, these studies and theories suggest that learning-by-giving-advice can be an effective tool for strengthening users' understanding. When a user is asked to provide advice, we display guidance above the text input field to encourage thoughtful, context-based responses. The \target~agent then receives the user's advice as a secondary input to inform its response to the \scammer~agent. 

\newt{Prior work on role-sensitive prompting and symbolic scaffolding shows that fixed-scenario structured prompts can stabilize LLM behavior and reduce contradictions in multi-party dialogue \cite{figueiredo2025symbolically}. Similarly, LLM-driven training systems and counseling chatbots demonstrate that agents can adapt to diverse user behaviors while still adhering to predefined conversational frames and goals \cite{meyer2025llm}. Therefore, to maintain coherent behavior even when participants offered unexpected or idiosyncratic advice, we carefully designed our prompts so that the scammer and target agents always operated within a fixed scenario and role schema. We additionally conducted a post-hoc analysis of how targets responded to user advice and discuss this in the Discussion (Sec. \ref{sec:discussion}).} %

\subsection{Component 4: Feedback and Summary} %
\newt{Finally, the last step of Kolb's ELT is reflecting on one's learning experience. Along these lines, prior work} has shown that in simulation-based training, immediate and structured feedback improves learning \cite{meir2024designing, gusukuma_2018_coding_feedback}. Hence, we devised a \feedback~agent that provides users with structured feedback based on their advice and the two agents' messages. Joakrim et al. found that between game-based training and context-based microtraining (CBMT), CBMT was more effective in teaching users to spot phishing, despite both improving performance \cite{kavrestad2022evaluation}. In another study, Hattie et al. found that feedback is a powerful tool in guiding students to the correct answer \cite{hattie_2007_power_feedback}. By \newtext{providing constructive} feedback, a user must think critically about their actions, which in turn can lead them to the correct response in the future. Therefore, the feedback agent is prompted (\img{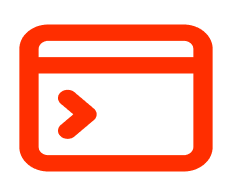}) to categorize users' actions as helpful or unhelpful, and this is shown to users. In addition to a binary rating, the feedback agent highlights how the user's advice impacted the target's behavior and the scammer's future responses. If the user’s advice was helpful, they see a congratulatory note summarizing what happened in that phase. Conversely, if the target succumbed to the scam, the user will see a summary informing them as such. The feedback component concludes with suggestions for improving their advice in subsequent phases and an overview of what to expect in the next phase.

\subsection{Implementation Details}
\textsf{ScamPilot} is a web-based application built using a React front-end UI and a back-end server built with Next.js and MongoDB, and hosted on Amazon AWS for reliability and load balancing. %
In theory, any interactive LLM API can be used to power \textsf{ScamPilot}'s roleplaying abilities \cite{tamoyan-etal-2025-llm}. Currently, we employ OpenAI's GPT-4o-mini \cite{openai_gpt4o_mini, openai2024gpt4technicalreport}. In our testing, we found that it produced text comparable to GPT-4 while being less prone to content flagging by OpenAI. After experimenting with several large language models (ChatGPT, Gemini, Claude), we qualitatively determined that GPT 4o-mini reliably reenacted both scammer and target agents’ personas and tactics. The \textsf{temperature} for the scammer system prompt is 0.5 and 1 for the target, with \textsf{max\_tokens} set to 8192. The web server retains all conversation history between the user and simulated interlocutors and handles API requests to three separate endpoints, one each for the scammer, target, and feedback agents.

\section{Methods}

\subsection{Experiment} %
To answer our research questions, we ran a between-subjects, randomized controlled experiment. Our experiment consisted of four interface conditions, each a variation of the \textsf{ScamPilot} system: (1) a naive \newtext{baseline} \textit{control}, (2) a \textit{quiz} interface, (3) an \textit{advice} interface, and (4) a \textit{quiz+advice} interface. 

\subsubsection{Participant Recruitment}
This study was approved by our institution’s IRB. We recruited participants from Prolific \cite{palan_prolificacsubject_2018}, an online survey platform. We restricted participation only to people located in the United States. We paid participants \$7 for 35 minutes at \$12/hour. %
To ensure sufficient statistical power and detect meaningful differences between experimental conditions \cite{yatani2016effect}, we ran a power analysis for four methods with $\alpha$ = 0.05, power of 0.8, and a medium effect size of 0.3, resulting in 32 participants per group.

\subsubsection{Experimental Conditions}
\label{sec:exptconditions}
Figure \ref{fig:experimentflow} provides an overview of our study design. Here we describe key interface differences between the four conditions. To answer the RQs and to ensure internal validity and reduce selection bias, participants were randomly assigned to one of the four conditions. All conditions consist of five dialogue turns for each of the three phases of the conversation. 

In the \textit{control} and \textit{quiz} conditions, conversations are pre-determined, with the former being about an irrelevant topic and the latter being about imposter scams. All participants in these two conditions see the same set of messages. At the end of each phase, participants are shown a pre-determined summary of the conversation. 

In the \textit{advice} and \textit{quiz+advice} conditions, conversations are dynamically generated through the STU pipeline, as discussed previously in Sec. \ref{sec:stupipeline}. In other words, no two participants saw exactly the same set of messages, although the content of the messages had significant overlap within each phase. Participants advance the conversation by either (a) providing advice to the target or (b) answering a quiz question correctly and providing advice to the target. At the end of each phase, participants are shown dynamically-generated feedback and summary messages relevant to their specific conversation. The key features of each condition are described as follows:

\begin{itemize}
\item \textbf{\textit{Control}}: Participants see a \newt{fixed}, pre-determined conversation between a mother and son about credit card skimming. \newt{This conversation was generated by two researchers and kept constant for all participants.} There are five dialogue turns (three from the mother and two from the son) per phase. Participants advance the conversation by pressing a \textsf{next} button. There are no multiple choice questions, or any mechanism for participants to provide input (advice) or receive feedback.
\item \textbf{\textit{Quiz}}: Participants see a \newt{fixed} conversation between the \scammer~and \target~agents engaged in dialogue. \newt{This conversation between the LLM-powered \target ~and \scammer ~agents was generated once and kept constant for all participants.} There are three messages generated by the scammer and two by the target per phase. Participants advance the conversation by correctly answering a quiz question.
\item \textbf{\textit{Advice}}: Participants engage with a dynamic, LLM-generated conversation between the \scammer~and \target~agents. \udot Users also provide advice to the target agent in real-time through the advice component.
\item \textbf{\textit{Quiz+Advice}}: Participants engage with a dynamic, LLM-generated conversation between the \scammer~and \target~agents. \udot Users must first correctly answer a multiple choice question before providing advice to the target agent.
\end{itemize}

\subsubsection{Procedure}

\begin{figure}[h]
    \centering
    \includegraphics[width=1\textwidth]{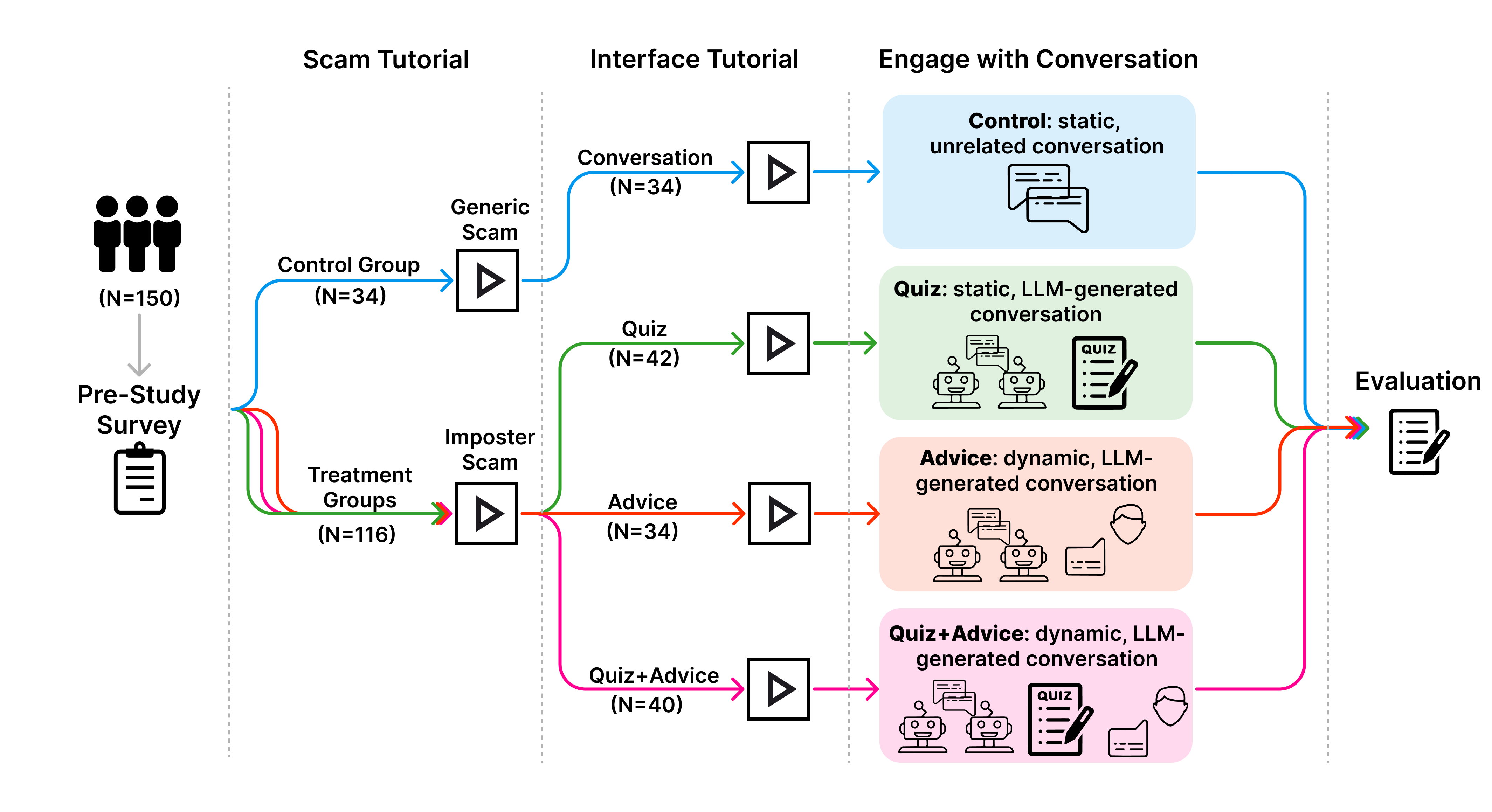}
    \caption{This figure shows the flow of the study for the 150 participants who completed the demographic and background survey as well as answered the attention checks correctly. Participants who failed the attention checks were excluded from this figure. Next, they were divided into one control and three treatment groups and assigned to each of the four interface conditions. All participants completed the same evaluation. N=17 did not pass all attention checks, thus we only report on the remaining N=150 participants.}
    \label{fig:experimentflow}
\end{figure}

Figure~\ref{fig:experimentflow} illustrates the overall study procedure. Participants first completed a pre-study survey (5 minutes), where they provided basic demographic information, described their prior experience with online scams, and responded to a series of survey instruments (described below). Then, depending on their assigned condition, participants watched a corresponding scam tutorial video and an interface tutorial video (6 mins. total). In the control group, the tutorial video was about credit card skimming. For the three treatment groups, the tutorial video described the characteristics of social engineering imposter scams. The interface tutorial video was tailored to each condition. Next, participants interacted with one of four \textsf{ScamPilot} interfaces (10 mins.), and completed a post-study evaluation survey (14 mins.).

\subsubsection{Measures}
To measure the effectiveness of the four interface conditions to promote learning, we leveraged a combination of existing and newly-designed survey measures. To understand participants' (change in) attitudes, we also adapted three of four sub-scales from Witte's Risk Behavior Diagnosis Scale \cite{witte1996predicting}: susceptibility to threat, self-efficacy, and response efficacy. In addition to participants' attitudes, we also measured participants' actual ability to detect a similar and different set of scams using a transfer learning framework for near and far transfer \cite{barnett_when_2002}. Near transfer refers to applying learned skills to new but structurally similar contexts, while far transfer involves applying those skills to a broader range of situations.

In the pre-study survey, we collected/measured participants': (1) demographic information, (2) prior exposure to scams, (3) security attitudes (SA-6 \cite{faklaris_self-report_2019}), (4) scam susceptibility \cite{james_correlates_2014}, and (5a) self-efficacy in fraud prevention \cite{chen_validation_2001}. In the post-study evaluation, we measured participants': (5b) self-efficacy in fraud prevention \cite{chen_validation_2001} (repeated measure), (6) response efficacy of the \textsf{ScamPilot} system \cite{chen_hacked_2020}, (7) scam/legitimate discernment ability, and (8) situational judgment ability:

\begin{enumerate}
    \item \textbf{Demographic Information}: We collected age, gender, occupation, and ethnicity. See Table \ref{table:demographics} in the Appendix.
    \item \newt{\textbf{Time Using the System/Post-Survey:} We collected the time participants spent on the system along with the post survey.}
    \item \textbf{Scam Exposure}: To determine a baseline and control for participants' prior experience, we also collected information about whether and how often participants had been exposed to scams, and how they were impacted. 
    \item \textbf{SA-6}: To further control for pre-existing differences between participants' privacy and security attitudes, we employed SA-6 \cite{faklaris_self-report_2019}, a six-item, five-point Likert scale, self-reported measure of security attitudes and is commonly used in cybersecurity surveys \cite{chen_hacked_2020, Albayram_2021_videoSA6, Krsek_2022_examiningSA6}. SA-6 has been empirically validated and strongly correlates with measures of self-reported recent security behavior, privacy, internet know-how, and perceived behavior control such as impulsivity.
    \item \textbf{Scam Susceptibility}: a three-item, five-point Likert scale measure adapted from prior studies \cite{james_correlates_2014, chung_reducing_2023} that measures the likelihood that a participant would fall for a threat. 
    \item \textbf{(Change in) Self-Efficacy}: a person's assessment of their ability to accomplish a goal (e.g., protect themselves from a scam) has been shown to be a strong predictor of behavior change \cite{RHEE2009816, sun2016}. We adapted four self-efficacy questions from the New General Self Efficacy scale \cite{chen_validation_2001}. To determine whether an experimental condition led to a \textit{change in self-efficacy}, we measure it twice: before and after using \textsf{ScamPilot}: \textit{SE1} and \textit{SE2}.
    \item \textbf{Response Efficacy}: a person's assessment as to whether a given action or tool will prevent a threat from occurring (here, being scammed). According to Protection Motivation Theory \cite{rogers_protection_1975}, a widely used framework for explaining protective behavior in the face of threats, people are more likely to take protective action when they believe the recommended response is effective. We included four response efficacy questions adapted from \citet{chen_hacked_2020} to measure participants’ perception of how well a given \textsf{ScamPilot} interface works and their experience using the system (RQ2). 
    \item \textbf{Scam/Legitimate Discernment}: To assess near-transfer learning, we measured participants’ ability to distinguish scam interactions from legitimate ones in unfamiliar but structurally similar conversational contexts. Each participant was presented with 12 screenshots of conversations on three platforms (iMessage, WhatsApp, Facebook Messenger) --- 6 scam and 6 legitimate. Each scenario is accompanied by two questions: likelihood of scam (7-point Likert-type from extremely unlikely to extremely likely) and a response justification (short free response).
    \item \textbf{Situational Judgment}: To evaluate far-transfer learning, we adapted situational judgment questions (SJQs) based on the Assessment of Situational Judgment \cite{getz_assessment_2024}. These realistic scenarios were more generic, but distinct from the ones participants had encountered thus far \newtext{in the study}. Each participant responded to 8 SJQs: 4 scam and 4 legitimate scenarios. For each scenario, participants rated their likelihood of compliance (7-point Likert-type from extremely unlikely to extremely likely).
\end{enumerate}

We additionally collected interaction and behavioral data through system logs. \newt{To control for external factors, we also} recorded the number of quiz attempts, the time participants spent on a quiz and giving each advice, the advice that participants gave, the entire conversation history and feedback, and responses to attention check questions. This system log data provided important insights into how participants engaged with the system and their decision-making process (RQ2 and RQ3). 

\subsubsection{\newt{Attention Checks}} \newt{We embedded six attention check questions throughout participants’ interaction with the system (see Table \ref{tab:attentioncheck} in the Appendix). This includes the pre-survey, after the tutorial video, and in the post-survey. We intentionally did not include any attention check questions during participants’ interaction in the chat component to avoid interrupting their learning process. Instead, we designed parallel attention-check items in both the pre- and post-surveys (e.g., If you are reading this, please select \textit{Agree}). Out of 167 total participants, 17 failed the attention check questions and were ultimately removed from our analysis.}

\subsubsection{\newt{Expected Outcomes}}
\newt{We designed our experiment and collected data to answer our three research questions. Although we did not have explicit hypotheses, we expected that the \textit{quiz+advice} interface would perform best across all research questions. We believed that the quiz questions would prompt participants to read the conversations more closely, enabling them to spot common tactics the scammer uses and, in turn, increase their scam discernment (RQ1). We also anticipated that having users perform a range of actions would increase their confidence in themselves and the system (RQ2, self- and response-efficacy), and that the quiz questions would guide users to provide action-oriented advice (RQ3).}

\subsubsection{\newt{Demographics}}
\newt{The demographics of our study participants as shown in Table \ref{table:demographics} in the Appendix are as follows. Of the 150 participants who passed the attention checks, 49.3\% identify as male, 48.0\% as female, 2.0\% as non-binary, and 0.7\% of participants preferred not to disclose their gender. Most participants are aged 25-34 (32.7\%), with other age groups represented as follows: 35-44 (22.7\%), 45-55 (21.3\%), 55-64 (11.3\%), 18-24 (8.0\%), and 65+ (4.0\%). The majority of participants are White (74.7\%), followed by Black or African American (16.0\%), Asian (6.7\%), Native Hawaiian/Other Pacific Islander (0.7\%), and 2.0\% of participants preferred to self-describe. There is a wide range of educational backgrounds with most participants holding a Bachelor’s degree (33.3\%), followed by high school or equivalent (29.3\%), Master’s degree (20.7\%), Associate degree (12.0\%), Doctorate degree (1.3\%), Professional degree (1.3\%), some formal education (0.7\%), and 1.3\% of participants self-described.}

\subsection{Data Analysis}
To comprehensively evaluate \textsf{ScamPilot}, we conducted a mixed-methods evaluation. This included: a quantitative analysis of performance measures and advice provided; and a qualitative analysis of advice provided.

\subsubsection{Quantitative Analysis}
We conducted quantitative analyses of the key variables collected in the post-study evaluation to understand the efficacy of our system and compare different learning outcomes across experimental conditions. 

To determine participants' ability to identify both scam and legitimate messages, we separated scam discernment and situational judgment measures into two separate scores each: \textit{scam} and \textit{legitimate}. Thus we calculated: (a) scam score (sum total; range:  -3 to +3), (b) legitimate score (sum total; range: -3 to +3), (c) situational judgment scam score (sum total; range: 1 to 7), (d) situational judgment legitimate score (sum total; range: 1 to 7), (e) change in self-efficacy (SE) score (SE2-SE1; range: 1-5), and (f) response efficacy score (sum total; range: 1-5). To obtain a given score, we sum totaled the corresponding numeric values for each sub-question, e.g., we converted a 7 point Likert-type scale to -3 to +3, and added the responses for each sub-question in that measure.

\textit{Statistical Tests.}
We used an ANCOVA model to produce Ordinary Least Squares (OLS) adjusted means, confidence intervals, and statistical analysis for \textit{interface conditions} while controlling for covariates. For sensitivity analysis, we use Iman-Conover non-parametric rank-based ANCOVA \cite{Iman_1982_nonparametric} because some of the residuals were skewed, which violates the normality of residual assumption for a standard ANCOVA. Rank transformations are common in HCI, such as the ART transformation \cite{Elkin_2021_ARTC}, but ART does not work with our data because we do not have a multi-factorial design, so the Iman-Conover non-parametric rank-based ANCOVA is more appropriate. The OLS adjusted means from a standard ANCOVA model are presented for better interpretability of results. OLS adjusted means do not rely on the ANCOVA assumptions, but the standard errors, confidence intervals, and significance from dependent variables with non-normal residuals can be biased and should be interpreted with caution. The rank-based ANCOVA is used for confirming statistical significance of interface condition since it is appropriate for data that violates the assumptions for standard ANCOVA, namely violation of normally distributed residuals. The ANCOVA model predicts different measured outcomes that were obtained through the post-study evaluation using four different interface conditions and controlling for important continuous covariates such as self-efficacy score, response efficacy score, SA-6 score, and total time on the system/evaluation. We assess\newt{ed} adherence/violation of standard ANCOVA model assumptions, which include linearity, normality of residuals, homogeneity of residuals, independence of residuals, and homogeneity of regression slopes. Most models met the homoscedasticity assumption (Breusch-Pagan test > 0.05), and for those that violated assumptions, we used a Heteroscedasticity Consistent 3 (HC3) correction which provides robust standard errors \cite{Long_2000_HC3} for the confidence intervals of the adjusted means. If we found a statistically significant F-test for the interface conditions factor in the standard ANCOVA model, we would perform post-hoc pairwise t-tests with a holm correction for multiple comparisons to identify which interface conditions differ. For example, if the overall interface condition F-test is significant, then we test which individual interface conditions are different. 

\subsubsection{Qualitative Analysis}
To complement our quantitative findings and gain deeper insight into participants’ reasoning strategies, we conducted a qualitative analysis of the advice given by participants to the target agent in the \textit{advice} and \textit{advice+quiz} conditions.

Adopting the iterative, team-based approach outlined by DeCuir-Gunby et al. \cite{Gunby2011}, our analysis involves three phases: codebook development, reliability assessment, and theoretical reflection. We then proceeded to content analysis \cite{krippendorff_sage_nodate} to quantify these patterns across experimental conditions.

The codebook development was an iterative, data-driven process: We developed a preliminary codebook based on pilot study data with 16 users where two authors open-coded these responses to identify recurring strategies. To further refine the initial codebook, the same two authors independently coded a random 10\% sample of the final dataset (N=444 free-text responses).

As recommended for team-based research \cite{Gunby2011}, we refined definitions and merged overlapping codes through discussion to better fit our research questions. This iterative process yielded a hierarchical codebook consisting of four distinct themes. Based on the advice's intent, we labeled these: \textit{theme 1}: advice encouraging verification, \textit{theme 2}: advice encouraging refusal, \textit{theme 3}: advice on managing the situation, and \textit{theme 4}: advice pointing out flaws. See Appendix Table \ref{tab:advicecodebook} for the full codebook.

\newt{To ensure consistency and objectivity in applying the final codebook, we proceeded with a rigorous reliability check.}
\newt{The two authors independently coded a fresh, randomly selected 10\% subset of the data. Since participants often provided lengthy, detailed advice containing multiple distinct strategies, our coding scheme allowed for
} 
multiple labels for a single response. 
\newt{Because of this, }we assessed IRR using Krippendorff’s Alpha \cite{krippendorff_sage_nodate} with a set-aware adaptation using a Jaccard distance metric \cite{alexander_jaccard_2016}. This approach is suitable for qualitative research where multiple codes can legitimately apply to a single unit \cite{passonneau-2006-measuring}. We achieved a Krippendorff’s Alpha of 0.87 ($>$0.80) indicating high agreement \cite{krippendorff_sage_nodate}. Finally, the two authors evenly divided the remaining 80\% of the data to code independently \cite{oconnor_intercoder_2020, mchugh_interrater_2012}.

\newt{To explicitly connect our analysis to design goals, we used the frequencies of these codes as dependent variables to support our theoretical concepts. In line with our design goal of providing an immersive experience, we operationalized RQ3 by using the relative frequency of theme 1 (advice encouraging verification) and theme 2 (advice encouraging refusal). This metric was chosen because decisive and relevant actions encouraged by participants serve as a robust indicator of immersion effectiveness. We expected that \textit{advice+quiz} compared to \textit{advice} condition would increase the prevalence of these specific, actionable codes.} \newt{Connecting back, as discussed in the related work section on Kolb's experiential learning theory, we focused on the stages of reflective observation and abstract conceptualization \cite{kolb_experiential_1984}. This is demonstrated as the participants' ability to explicitly identify the scam's mechanics or manage the emotional manipulation process. Therefore, we mapped this to theme 3 (managing the situation) and theme 4 (pointing out flaws). These codes represent the \newtext{users' sensemaking process of their} experience, which is different from the action-oriented feedback.} \newt{We then used Chi-square tests to compare the distribution of these code frequencies between conditions.}

\section{Findings}
\label{sec:quantfindings}
\begin{figure}
    \centering
    \includegraphics[width=0.75\linewidth]{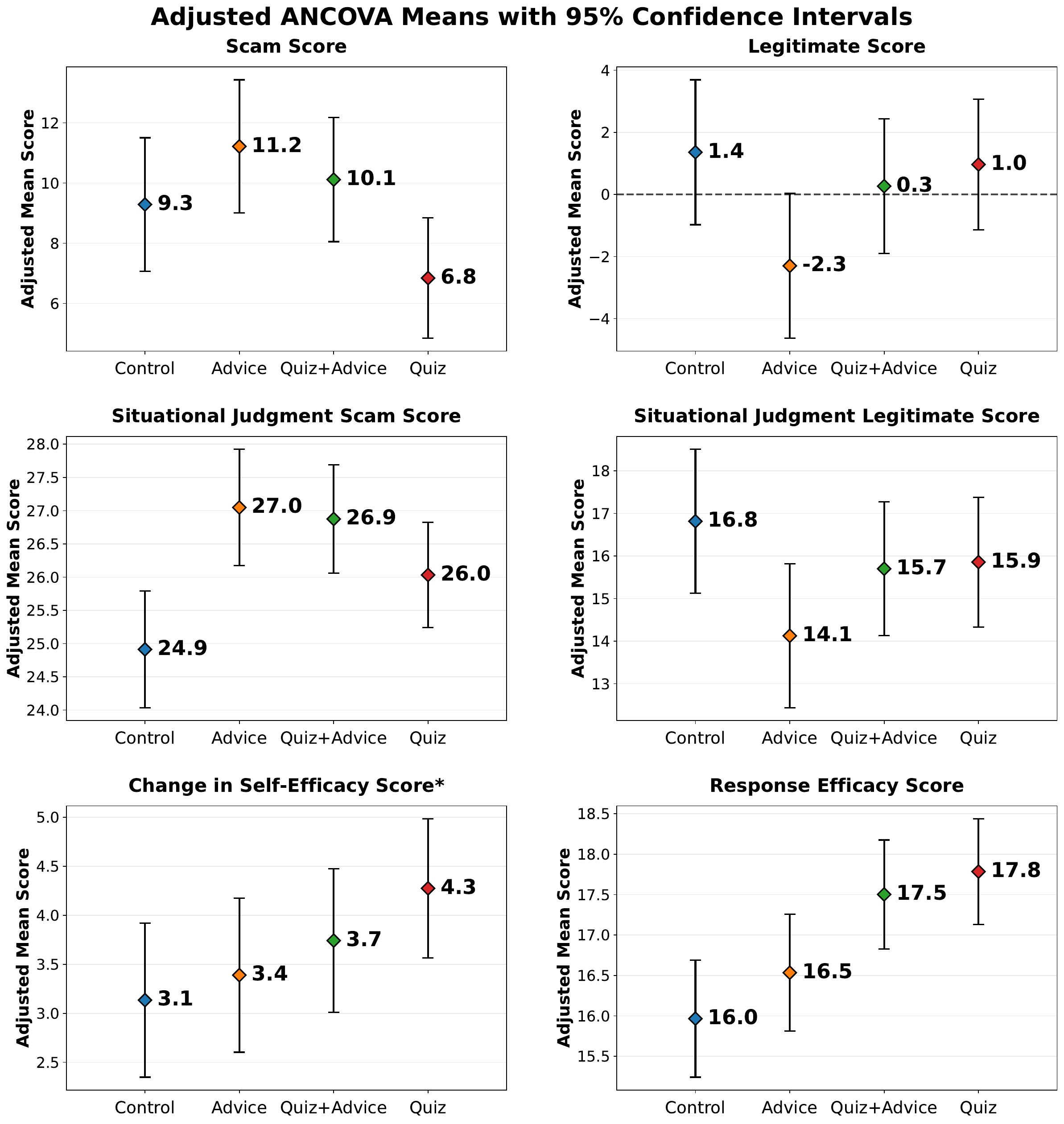}
    \caption{Error bars represent 95\% confidence intervals. All models \textit{Control} for SA-6 score, scam susceptibility score, self-efficacy score, completion time, and post-study survey time. * Self-efficacy score was excluded as a covariate in this analysis due to high correlation with the outcome variable.}
    \label{fig:ancova-adjusted-means-95ci}
\end{figure}

We first discuss our quantitative findings, which are summarized in Figure \ref{fig:ancova-adjusted-means-95ci} and Table \ref{tab:ancova_results}.

\subsection{\newt{Quantitative Findings For Determining the Best Interface for Scam Prevention (RQ1)}}
\newt{To answer RQ1, we discuss participants' performance on the following measures: scam discernment score, legitimate discernment score, situational judgment scam score, and situational judgment legitimate score.}

\begin{table}[H]
\begin{tabular}{llllll}
\hline
\rowcolor[HTML]{FFFFFF} 
{\color[HTML]{000000} \textbf{Dependent Variable}} & {\color[HTML]{000000} \textbf{F}} & {\color[HTML]{000000} \textbf{p-value}} & {\color[HTML]{000000} $\boldsymbol{\eta p^2}$} & {\color[HTML]{000000} \textbf{Effect Size}} & {\color[HTML]{000000} \textbf{Significant Covariates}} \\ \hline
\rowcolor[HTML]{EFEFEF} 
{\color[HTML]{000000} \textbf{Situational Judgment Scam Score}} & {\color[HTML]{000000} 11.41} & {\color[HTML]{000000} 0.0097*†} & {\color[HTML]{000000} 0.092} & {\color[HTML]{000000} Medium} & {\color[HTML]{000000} completion time, evaluation time} \\
\rowcolor[HTML]{FFFFFF} 
{\color[HTML]{000000} \textbf{Response Efficacy Score}} & {\color[HTML]{000000} 5.82} & {\color[HTML]{000000} \textbf{0.0009*}} & {\color[HTML]{000000} 0.110} & {\color[HTML]{000000} Medium} & {\color[HTML]{000000} self-efficacy score} \\
\rowcolor[HTML]{EFEFEF} 
{\color[HTML]{000000} \textbf{Scam Score}} & {\color[HTML]{000000} 3.04} & {\color[HTML]{000000} \textbf{0.0311*}} & {\color[HTML]{000000} 0.061} & {\color[HTML]{000000} Medium} & {\color[HTML]{000000} \begin{tabular}[c]{@{}l@{}}self-efficacy score, completion time, \\ evaluation time\end{tabular}} \\
\rowcolor[HTML]{FFFFFF} 
{\color[HTML]{000000} \textbf{Change in Self-Efficacy}} & {\color[HTML]{000000} 1.74} & {\color[HTML]{000000} 0.1622} & {\color[HTML]{000000} 0.035} & {\color[HTML]{000000} Small} & {\color[HTML]{000000} SA-6 score} \\
\rowcolor[HTML]{EFEFEF} 
{\color[HTML]{000000} \textbf{Legitimate Score}} & {\color[HTML]{000000} 1.94} & {\color[HTML]{000000} \textbf{0.1253}} & {\color[HTML]{000000} 0.040} & {\color[HTML]{000000} Small} & {\color[HTML]{000000} none} \\
\rowcolor[HTML]{FFFFFF} 
\textbf{Situational Judgment Legitimate  Score} & 5.77 & \textbf{0.1233†} & 0.034 & Small & SA-6 score, completion time \\ \hline
\end{tabular}
\caption{ANCOVA Results. Bold values indicate where the standard and non-parametric ANCOVA agree on significance. * indicate statistically significant p-values (p < 0.05). † indicate where the HC3 robust standard error correction was applied in the standard ANCOVA.}
\label{tab:ancova_results}
\end{table}

\subsubsection{Scam \newtext{Discernment} Score}
\textit{Interfaces with advice components increased scam recognition.} \newt{To answer RQ1, we examined the participants' ability to discern scam scenarios in near transfer situations.} The adjusted mean scores and their 95\% confidence intervals are 9.28 $\pm$ 2.22, 11.22 $\pm$ 2.21, 10.11 $\pm$ 2.06, and 6.84 $\pm$ 2.00 for \textit{control}, \textit{advice}, \textit{quiz+advice}, and \textit{quiz} interfaces, respectively. For the scam discernment questions, interface condition is a significant predictor of the scam score (F-test, p-value < 0.05) with a medium effect size (Partial eta-squared, $\eta$²p = 0.061). Interface condition in the Iman-Conover ANCOVA is also significant (F-test, p-value < 0.05) showing the standard ANCOVA result is supported. \newt{Since both tests are significant, this indicates that the type of interface  participants used had a significant impact on their ability to identify scams.} Users in the \textit{advice} condition are predicted to have a mean score 1.92 points higher than the \textit{control}, \textit{quiz+advice} are predicted to have a mean score 0.82 points above the \textit{control} and being in \textit{quiz} condition are predicted to have a mean score 2.44 points lower than the \textit{control} condition. Looking at the other covariates, the self-efficacy, total completion time and post-study survey time are all significant predictors (F-test, p-values < 0.05) and also supported by the Iman-Conover ANCOVA test (F-test, p-value < 0.05). With higher self-efficacy score and longer time on the post-study survey increasing the scam score and the total time for users to complete the system lowering their scam score. Since the interface condition is significant, in the post-hoc pairwise t-tests we find that the \textit{advice} condition was significantly higher (+4.37 points) than the \textit{quiz} condition (t-test, holm-corrected p-value <0.05). The \textit{quiz+advice} condition is significantly higher (+3.27 points) than \textit{quiz} before the holm correction but after the correction it is not (t-test, holm p-value > 0.05, p-value <0.05)\newt{, indicating that the advice component led to greater performance on scam discernment questions}. The other interface conditions are not significantly different (t-test, holm-corrected p-values > 0.05). 
\newt{Thus, participants who engaged with the advice component achieved greater learning on near transfer tasks, leading to increased scam discernment}.

\subsubsection{Legitimate \newtext{Discernment} Score}
\textit{Interfaces with advice components did not do significantly worse at identifying legitimate conversations.} \newt{To address RQ1, we evaluated how effectively participants from each interface condition can discern legitimate conversations in near transfer scenarios.} The adjusted mean scores and the range of their 95\% confidence interval for these questions are 1.35 $\pm$ 2.33, -2.30 $\pm$ 2.33, 0.26 $\pm$ 2.17, and 0.96 $\pm$ 2.10 for \textit{control}, \textit{advice}, \textit{quiz+advice}, and \textit{quiz} interfaces, respectively. For the scam/legitimate discernment questions, the interface condition is not a significant predictor of the legitimate score (F-test, p-value > 0.05) for a small effect size (Partial eta-squared, $\eta$²p = 0.040) which is confirmed by the Iman-Conover ANCOVA. This shows that none of the experimental conditions caused significantly worse scores on scam discernment questions. \newt{Thus, these results demonstrate that the advice conditions are the most effective interfaces, as they significantly increase scam discernment scores (above) \textit{without} significantly reducing performance on legitimate conversations.}

\subsubsection{Situational Judgment Scam Score}
\textit{Interfaces with advice components perform better than the control on far transfer questions}. \newt{To continue answering RQ1, we analyzed participants' ability to distinguish potentially suspicious generic scenarios.} The situational judgment scam score fails the homoscedasticity and normality of residuals assumptions. Since it fails homoscedasticity, we use the HC3 correction. To account for this we use the Iman-Conover ANCOVA as a sensitivity analysis and only the standard ANCOVA's means can be interpreted without caution. The adjusted means and the range for the 95\% confidence interval for the situational judgment scam score are 24.91 $\pm$ 0.88, 27.05 $\pm$ 0.88, 26.88 $\pm$ 0.82, and 26.03 $\pm$ 0.79 for \textit{control}, \textit{advice}, \textit{quiz+advice}, and \textit{quiz} conditions, respectively. Looking at these intervals, the \textit{advice} and \textit{quiz+advice} have no overlap with the \textit{control}. Users in the \textit{advice} condition are predicted to score 2.13 points higher than the \textit{control}, \textit{quiz+advice} condition are predicted to score 1.96 points higher and \textit{quiz} is predicted to score 1.12 points higher than the \textit{control}. \newt{These intervals indicate that the control group scored lower than all other conditions, with the advice conditions achieving the highest scores. This suggests that participants in the advice conditions were more likely to identify these scam scenarios.} For the situational judgment questions, interface condition is significant for the standard ANCOVA (F-test, p-value < 0.05) with a medium effect size (Partial eta-squared, $\eta$²p = 0.092) but is not significant for the Iman-Conover ANCOVA (F-test, p-value = 0.058) while it is trending. For the post-hoc testing, the \textit{advice} and \textit{quiz+advice} interface conditions were both significantly higher than the \textit{control} for the situational judgment scam score questions (t-test, p-value < 0.05, holm p-value <0.05). \newt{These results show that the participants who used the advice interface performed significantly better at identifying far transfer scenarios than the control, which aligns with the confidence intervals.} Since our data fails these assumptions, people should not overinterpret the significance tests, standard errors, and confidence intervals. Since the \textit{quiz+advice} and \textit{advice} interfaces have the highest means, this indicates they do the best on these questions overall and so this helps answer which interface is best\newt{, while confirming that our experimental conditions performed the best on scam questions}.

\subsubsection{Situational Judgment Legitimate Score}
\textit{The interfaces with advice components do not perform significantly worse on the far transfer legitimate questions.} \newt{To answer RQ1 and understand how well participants could decipher legitimate situations, we analyzed their situational judgment legitimate score.} Since this model failed the homoscedasticity test, we used a HC3 corrected ANCOVA model. The adjusted means and the range for the 95\% confidence interval for these questions are 16.82 $\pm$ 1.69, 14.13 $\pm$ 1.69, 15.7 $\pm$ 1.57, and 15.86 $\pm$ 1.52 for \textit{control}, \textit{advice}, \textit{quiz+advice}, and \textit{quiz} conditions, respectively. For the situational judgment questions, the interface condition is not a significant predictor of the situational judgment scam score (F-test, p-value > 0.05) which is confirmed by the Iman-Conover ANCOVA (F-test, p-value > 0.05) for a small effect size (Partial eta-squared, $\eta$²p = 0.0344). \newt{These tests show us that the \textit{advice} and \textit{quiz+advice} conditions do not perform significantly worse on discerning legitimate situations.} Looking at the other covariates the total completion time and SA-6 score are significant predictors (F-test, p-value < 0.05). A higher SA-6 score decreases the situational judgment legitimate score and the total time for users to complete the system lowering their situational judgment legitimate score. None of the other variables are significant either (F-test, p-value > 0.05). The fact that the experimental conditions \textit{do not} significantly lower legitimate scores is important for understanding which interface performs best\newt{. These findings support that our experimental condition, \textit{quiz+advice}, performs similarly to other interfaces in legitimate situations.} 

\subsection{\newt{Quantitative Findings For Determining the Best Interface for Response and Self-Efficacy (RQ2)}}
\newt{To answer RQ2, we analyze the change in self-efficacy and response efficacy scores. This analysis reveals how users' confidence levels changed and how they perceive the system's helpfulness.}

\subsubsection{Change in Self-Efficacy Score}
\textit{The interfaces with \textit{quiz} components lead to a greater change in self-efficacy score.} \newt{To answer RQ2, we analyzed users' change in self-efficacy before and after using our interface conditions.} The change in self-efficacy score is measured by taking the difference between the same set of self-efficacy questions that participants are asked before and after using the system. This dataset has a skew in the normality of the residuals so these results need to be interpreted with caution. The adjusted means and the range for the 95\% confidence interval for these questions are 3.13 $\pm$ 0.79, 3.39 $\pm$ 0.79, 3.74 $\pm$ 0.73, and 4.27 $\pm$ 0.71 for \textit{control}, \textit{advice}, \textit{quiz+advice}, and \textit{quiz} conditions, respectively. For the self-efficacy questions, interface condition is not a significant predictor of the change in self-efficacy score for the standard ANCOVA (F-test, p-value > 0.05) with a small effect size (Partial eta-squared, $\eta$²p = 0.035) but the Iman-Conover ANCOVA test is significant for interface condition (F-test, p-value < 0.05). The difference between these significance tests may be due to the skew in the data which the Iman-Conover test deals with better than a standard ANCOVA. \newt{While we did not observe a significant difference between conditions, self-efficacy scores increased in every group, indicating that interventions generally lead to participants feeling more confident.} SA-6 score is a significant covariate (F-test, p-value <0.05) and people with higher SA-6 scores have a lower change in their self-efficacy score. \newt{Even though every group displayed an increase in self efficacy, the \textit{quiz} and \textit{quiz+advice} conditions had the greatest increase, suggesting that those who used the embedded quiz questions perceived a greater increase in self-efficacy.}

\subsubsection{Response Efficacy Score}
\textit{Interfaces with a quiz component lead to a higher response efficacy score.} \newt{For RQ2, we also evaluated how helpful participants found our different interface conditions. Thus, we analyzed their change in response efficacy scores.} The adjusted means and the range for the 95\% confidence interval for these questions are 15.97 $\pm$ 0.72, 16.53 $\pm$ 0.72, 17.5 $\pm$ 0.67, and 17.78 $\pm$ 0.65 for \textit{control}, \textit{advice}, \textit{quiz+advice}, and \textit{quiz} conditions, respectively. For the response efficacy questions, the interface condition is a significant predictor of response efficacy score (F-test, p-value < 0.05) and the Iman-Conover ANCOVA test is also significant for interface condition (F-test, p-value < 0.05) for a medium effect size (Partial eta-squared, $\eta$²p = 0.11). \newt{These results indicate that the interface condition used by the participants significantly affected their change in perceived response efficacy.} Users in the \textit{advice} condition are modeled to score 0.55 points higher than the \textit{control}, users in \textit{quiz+advice} are predicted to score 1.53 points higher than the \textit{control} and users in the \textit{quiz} condition are predicted to score 1.81 points higher than the \textit{control}. Looking at the other variables, self-efficacy score is also a significant predictor (F-test, p-value < 0.05). With a higher self-efficacy score there is an increase in response efficacy score. In the post-hoc pairwise test, we find that the \textit{quiz} (+1.82 points) and the \textit{quiz+advice} (+1.54 points) conditions are significantly higher than the \textit{control} condition (t-test, holm p-value <0.05). The \textit{quiz} condition is also significantly higher (+1.25 points) than the \textit{advice} interface before the holm correction but is not significant after correction (t-test, p-value < 0.05, holm p-value >0.05). \newt{In other words, the interfaces with a quiz component scored significantly higher than the control, indicating that participants felt much more confident in the system’s effectiveness when a quiz was included.} None of the other conditions are significantly different from each other (t-test, holm p-value > 0.05). This shows that the users who have embedded quiz questions in their interface believe the system is more effective.

\subsection{\newt{Qualitative Analysis of Types of Advice Given By the Participants (RQ3)}}

\newt{To address RQ3, we analyzed the advice content across all conditions. This qualitative analysis identified four primary themes, ranging from concrete actions such as verification and refusal to cognitive strategies like emotional management and logical analysis. \newtext{While there were differences in the quantitative findings for the \textit{quiz+advice} and the \textit{advice} interfaces, they were not significant, and therefore the performance for the different types of advice is similar across the interfaces. This shows that the interface mattered more for performance than the type of advice.}}

\subsubsection{Theme 1: Advice Encouraging 
\newt{Verification}}
We found that participants provided advice encouraging the target to take concrete actions 
\newt{of verification}. 56.8\% of all advice matches at least one code with this theme. We found three main verification strategies: (i) direct verification of the sender’s identity, (ii) verifying or consulting through trusted sources, and (iii) confirming the plausibility of the situation. For example, \newt{direct identity verification is being used here:}
\textit{``Ask him the name of your pet. If it is Mark he should be able to answer correctly.''} 
Another participant suggested a different strategy to \textit{``call Mark's dad to verify without telling him Mark's in trouble''}. Here, the focus shifts to checking with trusted sources, as the target is advised to reach out to close relatives to confirm if the situation is legitimate. As another participant suggested, \textit{``ask him to tell you his precise location.''} This advice focuses on collecting more information and 
\newt{verifying} the plausibility of the situation.

\subsubsection{Theme 2: Advice \newt{Encouraging Refusal}}
 Another major theme centered on 
certain actions \newt{of refusal}. 18.5\% of all advice matches at least one code with this theme. Advice here focused on refusal strategies related to information sharing and communication: (i) not sharing personal information, (ii) not transferring money, (iii) not engaging further in the conversation, and (iv) refusing until a condition is satisfied. For example, one participant warned, \textit{``Don't send money or personal information, try to get information from Mark.''} This participant explicitly advised the target to limit information sharing. Another participant suggested \textit{``Tell them you need to speak with them directly or else you can't help. Tell them that you won't discuss it over text any longer.''} This example shows that participants also used the strategy of refusal under condition, advising the target to continue responding until a condition was satisfied.

Taken together, these comments demonstrate strategies of refusal. Whether by blocking information sharing, cutting off transactions, disengaging entirely, or placing conditions on compliance, participants framed ``saying no'' as an important counterpart to proactive verification.

\subsubsection{Theme 3: 
\newt{Advice on Managing the Situation}}
\newt{One theme of }participant's advice
focused on process-oriented feedback, emphasizing how the target should emotionally and cognitively manage the situation. 9.5\% of all advice matches at least one code with this theme. Three recurring strategies were identified: (i) recognizing emotional manipulation, (ii) providing emotional support, and (iii) urging caution. For example, one participant stated, \textit{``If what the scammer is saying is true, then you would be breaking the trust of your grandson’s parents! They are using this sense of urgency so you do not find out that this is not really Mark.''} This participant identified the scammer’s use of urgency as a tactic to pressure the target into acting without verification, and also discouraged them from consulting other trusted sources. Another participant offered emotional support, noting, \textit{``Try not to get emotional about this.''} Similarly, one participant urged caution, warning, \textit{``Be careful, the scammers will impersonate relatives in a crisis. Wait, do not be in a hurry --- find out whether it actually IS Mark.''} This shows that users can identify how scammers may impersonate relatives in urgent situations and advised the target to slow down and verify identity before responding. These examples show that instead of directing concrete steps such as identify verification or refusing action, process-related advice emphasized maintaining composure, awareness, and critical judgment. 

\subsubsection{Theme 4: Advice 
\newt{Pointing Out Flaws}}
Lastly, participants actively sought to undermine the scammer’s credibility by providing additional information or logical reasoning\newt{, pointing out the flaws in the conversation}. 15\% of all advice matches at least one code with this theme. Instead of recommending direct verification or outright refusal, participants: (i) identified contradictions, (ii) pointed out vague or generic claims, or (iii) proposed alternative explanations to challenge the scammer’s story. For example, one participant noted, \textit{``That seems like a very generalized statement and doesn’t give you any details that it is in fact Mark.''} This reflects the strategy of pointing out vague or generic claims, where participants highlighted the lack of personalized details as a warning sign that the message could have been sent to anyone. Another participant explained, \textit{``I think it's very suspicious that he wants to keep it from his parents. He is also being very pushy with asking for help immediately.''} Here, the participant pointed out that secrecy from family and exaggerated urgency were inconsistent with how a real grandson would act\newt{. This} 
was \newt{also} emphasized in the training. A third participant used external evidence to identify logical inconsistencies in the scammer's story: \textit{``Mark just said \$200 and then \$300, this is suspicious. Also, tickets in Boston don't cost this much.''} %

\subsection{\newt{Quantitative Differences in Advice Between Conditions (RQ3)}} 

\newt{To evaluate the difference in outcome of the conditions, we examined quantitative differences in the verbosity and strategic content of the advice provided by participants.} \newt{Before examining the content of the advice, we analyzed the word length to measure participants' engagement.} \newt{Statistics indicated that advice in the \textit{advice} condition (Mean = 25.77, SD = 26.90, Median = 19.00, n = 204) was longer than in the \textit{quiz+advice} condition (Mean = 18.36, SD = 10.36, Median = 16.00, n = 240). A Mann-Whitney U test revealed a significant difference in the distribution of advice word length between the two conditions (U = 28353.50, p = .003), with a medium effect size (Cohen's d = 0.374, r = 0.141). A one-tailed test confirmed that advice length was significantly greater in the \textit{advice} condition compared to the \textit{quiz+advice} condition (p = .001.)}

\newt{This difference in verbosity aligns with our design goal of experiential learning. The shorter \textit{quiz+advice} responses suggest that the interactive quiz successfully \textit{shifted} participants more towards the active application stage in Kolb's Experiential Learning Theory \cite{kolb_experiential_1984}, where they can test and apply their newly formed concepts in real-world situations. Having just been immersed in a scam scenario, participants in the \textit{quiz+advice} condition could provide more directed and decisive advice without needing to elaborate. Conversely, the longer and more variable responses in the \textit{advice} condition suggests that participants likely felt a greater need to verbally reason through each scenario, providing longer explanations alongside their advice.}

\subsubsection{Theme-level differences.}  
We found that overall, the distribution of advice categories between \textit{advice} and \textit{quiz+advice} was significantly different ($\chi^2$(3)=8.10, $p=0.044$, Cramér’s V=0.118). \newt{Mapping these results, we found a notable difference in the learning stages these participants prioritized in the two conditions.} 

\newt{\textbf{\textit{Advice} participants focused on reflective observation:} Participants in the \textit{advice} condition leaned towards strategies associated with reflective observation and abstract conceptualization stages of experiential learning.} For example, \newt{advice on managing the situation} 
were more frequent in the \textit{advice} condition (12\% vs. 7\%).
\newt{These responses often focused on cognitive processing, such as }\textit{``That's good, keep up the good work of not sending anything before you have verification. I know this person sounds like they are in trouble, but they could be manipulating you.''} Similarly, \newt{advice pointing out flaws} 
(17\% in \textit{advice} vs. 13\%) tended to highlight inconsistencies without direct refusal. \newt{These differences suggest that without the quiz component, participants engaged in cognitive aspects of learning as they analyzed the scam's mechanics and emotions, but they were less likely to suggest decisive actions.}

\newt{\textbf{\textit{Quiz+advice} participants focused on active experimentation under immersion}:} Conversely, participants in the \textit{quiz+advice} condition more often gave action-oriented responses. \newt{They often provided concrete directions.}
\newt{Specifically, advice encouraging refusal} (21\% vs. 15\%), such as \textit{``Tell him you don't feel at all comfortable about sending money electronically.''} \newt{Advice encouraging verification also has a slight increase} 
(58\% in \textit{quiz+advice} vs. 55\%), such as \textit{``Tell him to take a selfie of himself and the background.''} While these shifts did not reach statistical significance after correcting for multiple comparisons, they demonstrate a pattern in which the \textit{advice} condition elicited more recognition-based comments, whereas the \textit{quiz+advice} condition elicited more guidance on taking specific actions. 

\subsubsection{Code-level differences.}  
We first examined the overall association between experimental condition and the distribution of advice codes, which yielded a significant omnibus chi-square test ($\chi^2$=30.09, $df$=18, $p$=0.037). To identify which specific codes contributed to this effect, we conducted a series of post-hoc 2×2 tests comparing the relative frequency of each code between the two conditions. After adjusting for multiple comparisons using the Bonferroni correction, only the \textit{refuse to take action} code remained significant (Bonferroni-adjusted $p$=0.0097). This indicates that the overall distributional difference was primarily driven by participants in the \textit{quiz+advice} condition being \textit{more likely} to produce explicit refusals, such as \textit{``Do not give him anything! Cut him off.''} \newt{Explicit refusal is a socially difficult action that requires confidence, and the fact that \textit{quiz+advice} participants were more likely to provide such ``hard stop'' advice suggests that our intervention helped participants be more confident --- which contrasts with our finding that participants in the \textit{quiz} condition had higher self- and response efficacy scores (Sec. \ref{sec:quantfindings}).} All other codes showed no reliable differences after correction, indicating that while minor shifts were observed, they were not significant after adjustment.

\section{Discussion}
\label{sec:discussion}

\newt{Our findings suggest that inoculation theory and experiential learning, combined with giving advice, is an effective method for scam education. We demonstrate that the \textit{quiz+advice} method is the most effective interface across all performance metrics. Participants in this method perform better than the \textit{control} group on the scam discernment questions (+8\%), scam situational judgment questions (+8\%), change in self-efficacy score (+19\%), and response efficacy score (+9\%), while only scoring approximately 1 point lower on legitimate discernment and situational judgment questions. This is common and expected for scam education, as teaching people about these scams will likely increase their concern about scams \cite{johnson_inoculation_nodate, Modirrousta_2023_innoculation_skepticism}. However, through our design, we minimized this negative effect, resulting in the legitimate questions not differing significantly for the \textit{quiz+advice} condition compared to the \textit{control}. This outcome aligns with findings for a gold standard inoculation system \cite{chowhery_2023_QA}. The \textit{quiz+advice} version of ScamPilot is a valuable tool for scam education.}

\newt{In the remainder of} this section, we discuss the implications of the different interfaces, how inoculation theory works for scam education, implications for cybersecurity interventions, design implications for conversational user interfaces (CUI), and ethical considerations, limitations, and future work. 

\subsection{Quizzing Combined with Advice Enhances Learning}
To answer RQ1, we showed that the \textit{quiz+advice} interface is the most effective for learning and scam detection. Misidentifying a \textit{legitimate message as a scam} is less of a concern than wrongly identifying a \textit{scam as legitimate}. Ideally, participants should score higher by correctly identifying scam questions while not lowering their legitimate scores. The \textit{quiz+advice} method is the interface that best accomplishes this goal: it balances increased sensitivity towards scam recognition against increased wariness of legitimate messages as potential scams.

Our study also showed that when participants provided advice to the target agent, they performed better on the scam discernment questions. While the advice conditions performed worse on legitimate questions, the difference was not statistically significant. The advice methods utilized the ``learning-by-teaching'' principle \cite{allen_learning_1973, kolb_experiential_1984, bargh_cognitive_1980, benware_quality_1984} and our results showed that this framework is effective for teaching people common scammer strategies. Being asked to provide advice leads participants to engage more with the conversation and therefore participants are more likely to pick up on red flags. However, this increased engagement also made participants more wary of the legitimate questions, similar to other scam education training and games \cite{johnson_inoculation_nodate, roozenbeek_psychological_2022}. \newt{To answer RQ3,} our qualitative analysis of advice supports this interpretation: participants in the \textit{quiz+advice} condition were more likely to produce actionable recommendations, while \textit{advice} participants more often provided process-related advice or exhibited caution. This shows that the quiz alongside advice effectively prepared and directed participants towards more concrete, actionable advice
that is useful in practice. 

\newt{Because our system divides a scam into three phases, it is possible that advice given by a user in one phase may not be relevant in a subsequent phase. We thus reviewed all of the advice given by participants, and found that while some advice was strategically ineffective (e.g., advising the target to comply), no advice was contextually inappropriate or contradictory. This consistency enables us to evaluate our findings without the interference of confounding variables, such as unpredictable dialogue shifts. Furthermore, to ensure that the LLM conversations did not introduce any unsafe elements that might have deviated from the initial prompt's intended purpose and to maintain data quality and research validity, we systematically screened every single dialogue after the full study. We did not identify any conversation that could be considered potentially harmful \cite{Ji_2023}.} 

We also found that even though participants with quiz questions had a 6\% higher change in response efficacy and 8\% higher change in self-efficacy scores than the next closest advice interface, the participants performed \textit{worse} on legitimate discernment questions. Participants in the \textit{quiz} condition also performed worse on both the scam and legitimate discernment questions, even though it had the highest change in response and self-efficacy scores. In other words, \newt{and also answering RQ2}, participants in the \textit{quiz} condition \textit{perceived} an increase in their ability to detect scams but without an accompanied increase in actual performance. Surprisingly, participants also rated the \textit{quiz} condition the highest in terms of response efficacy. Consequently, participants with embedded quiz questions tended to have a higher perception of their abilities for scam discernment \cite{chen_hacked_2020, van2021effects} while also (incorrectly) strengthening their belief that quizzes alone are an effective education mechanism \cite{chen_hacked_2020}
. 
\newt{Overall, however, we find that} the \textit{quiz+advice} method blended the strengths of both the \textit{quiz} and \textit{advice} interfaces. Participants had higher scores on the scam questions, higher change in response and self-efficacy score while lowering the reduction in legitimate questions to 1 point. Thus, the \textit{quiz+advice} method is effectively using the PTPA model \cite{workman_study_2022} to teach scam recognition, leading to participants scoring higher on scam questions \textit{without} doing significantly worse discerning legitimate questions.

\subsection{Implications of Inoculation Theory with LLMs}
Prior work showed that inoculation theory works well for information manipulation tactics online and for email scams \cite{roozenbeek_psychological_2022, robb_who_2023} and due to this we hypothesized that inoculation theory would be effective for identifying scam conversations. Our study provides additional evidence that inoculation theory is applicable to scams, where participants build the knowledge of how scam conversations progress and the associated red flags. The participants who had dynamic LLM-generated conversations overall performed better on the scam discernment questions and situational judgment questions compared to the interfaces that had static conversations, showing that participants could recognize scam conversations at a higher rate. The \textit{quiz} interface, which has a scam conversation but is not dynamic, did worse on the scam discernment questions than the other three interfaces, but did perform better on the situational judgment questions than the \textit{control}. Thus, the interfaces with dynamic LLM-generated conversations, \textit{advice} and \textit{quiz+advice}, performed better at scam discernment than the \textit{control} and \textit{quiz} with static scam conversations. This shows that inoculation theory can be effective at allowing participants to not only understand scams but to also be able to better identify scam conversations. \newt{Therefore, the interfaces utilizing inoculation theory perform best, supporting RQ1}. 

Participants with the dynamic LLM-generated conversations (\textit{advice} and \textit{quiz+advice} conditions) tended to identify legitimate situations as scams for the discernment questions at the two highest rates. Comparing this to the situational judgment questions, the participants with dynamic conversations were not much worse at identifying legitimate situations. We propose that this is the effect of people seeing scam conversations in a messaging interface \textit{and then} being tested on conversations in this same format --- leading to more false identifications of scams. Since the situational judgment questions did not have the same format, this was not an issue and it supports this theory. 

The LLMs provide an interactive experience that engages participants and makes them feel like they are involved in the scam conversation. While the dynamic conversation accomplished this, with the lower comparative score for the \textit{quiz} interface, we hypothesize that by making the participants feel like their decisions affected the target agent, they were more inclined to pay attention to nuanced details and point out any red flags and inconsistencies in the scammer agent’s messages. 

Similar to the codebook for advice, we designed one for justifications. However, we omit it from our Findings because we found minimal differences between conditions. This suggests that while different interface conditions influenced scam recognition performance and the advice that participants gave, they did not seem to affect participants' justifications as to why a conversation was a scam or legitimate.

Our findings show that using inoculation with conversations generated by LLMs can be expanded to other scam situations. 
We believe that ScamPilot system would perform equally well with other types of scams, \newt{as most scams can be broken down into three phases \cite{oak_hello_2025}} \textsf{ScamPilot} only needs slight modifications to system prompts and the multiple choice questions changed to support other scam types, such as tech support scams, IRS scams, pig butchering scams \cite{oak_hello_2025}, etc. 

\subsection{Implications for Cybersecurity Interventions}
The \textsf{ScamPilot} system is dynamic and allows for users to interact with it in multiple ways, which we claim is essential for new cybersecurity interventions. Scams have gotten so advanced that they are usually not targeting infrastructures but instead humans through tailored socially engineered attacks \cite{jeong_towards_2019, rahman_human_2021}. Cybersecurity applications must keep up with the different ways that cybercriminals are creating social engineering attacks. However, there is a difference between hearing and learning abstractly about attacks versus directly experiencing an attack \cite{jensen2024awareness, bada2019cybersecurityawarenesscampaigns}.

The interaction techniques from \textsf{ScamPilot} can also be expanded to other cybersecurity applications. %
\newt{One area where this \textsf{ScamPilot}'s interaction technique can be used is in workplace training. Many organizations require employees to undergo onboarding or periodic cybersecurity training \cite{Katsarakes_2024_onboarding, WONG_2022_awarness}, which often include scam-prevention or phishing training, but with mixed results \cite{Tally_2023_officeScams}. Our results point to a potentially new interaction modality in cybersecurity trainings by demonstrating the tactics and behaviors commonly used by attackers. For example, workplaces could identify the three to five most prevalent attacks that they face, and could then create tailored prompts for each scenario. Employees would be able to review each type of attack in a reasonable amount of time, exposing them to the attack their employers are most concerned about and helping them recognize the red flags associated with them. A key challenge is keeping the attack scenario prompts up to date.}

\subsection{Implications for the Design of Conversational User Interfaces for Learning}
While our system focuses on scam recognition, we suggest that having dynamically generated conversations with LLMs while allowing for user input could be used in other domains. This idea of simulating a realistic conversation while asking for user input makes it so that participants are able to interact with a conversation while not being directly part of the conversation. By giving an avenue for people to change the course of a conversation and the content in the conversation, the impacts of decisions are shown in real time, whether positive or negative. Allowing participants to see the real time consequences of their advice provides instant feedback and this can be used for situations like generating stories, playing out conversations between friends, or educational purposes \cite{shaikh2024rehearsal}. 

With advances in LLMs and their ability to create personalized stories and conversations, new CUIs for learning should focus on personalized aspect of learning. Making the user feel that they are a part of a conversation allows them to feel more immersed. \newt{Our finding in response to RQ3 supports this, as participants consistently provided detailed and lengthy advice in the two advice conditions. This suggests that the participants were highly engaged and enjoyed the conversational format. This commitment of effort and time suggests a positive user experience, and shows that learning through personalized CUIs can be motivating.} Providing advice as a form of interactive learning is something that CUIs should look to incorporate in future research projects. This can be applied in systems where personalization and real-time feedback matter such as educational discussions of literature, collaborative learning environments, and idea creation. %

\subsection{Ethical Considerations, Limitations, and Future Work}
\newt{Deploying \textsf{TargetPractice} in the wild presents a number of ethical and legal concerns.} The fact that we are able to use commercially-available LLMs to simulate scam conversations through prompt engineering techniques is a cause for concern that should be addressed. %
\newt{However, not all LLM vendors may allow such a (mis)use in the future or at a large scale and may block such usage. Vendors may also change their terms of service or retune their models at a whim, making it difficult to ensure a consistent user experience over time. Our interface also required some effort to prompt engineer a commercially available LLM, and scaling and customizing a version for a specific company or use case would require additional effort.}

Along these lines, although our study found that showing scam conversations can inoculate users, we have not \newt{yet evaluated our approach in real-world situations and with more diverse users} to see how the simulated training can be transferred to real-world situations. Additionally, experiencing a scam can be traumatic \cite{oak_hello_2025, Balcombe_2025_scam_trauma} and thus our goal was to minimize this factor in the design of \textsf{ScamPilot}. To minimize the trauma caused by experiencing a scam, we place the user in an advisory capacity and not in the position of the person being scammed. We also do not depict the target losing money or being emotionally distraught. Through these two decisions, we sought to mitigate the potentially traumatizing aspects of scams. \newt{However, we acknowledge that we did not formally evaluate the psychological impact of the system in our current study. In future work, we will focus on developing and implementing safety evaluations and also measure the emotional and psychological experience of the participants during training.} 

The first limitation of our work is we were only able to evaluate imposter scams. While we believe that our findings extend to other scam types, we do not yet have empirical evidence. Another limitation is that scams are ever-evolving, requiring our system to be constantly updated. Modifying system prompts for \textsf{ScamPilot} to showcase new scammer tactics requires minimal effort due to the use of few-shot prompting.

A limitation of our findings is that we do not test for changes in user’s ability (within-subject) but for differences in groups (between subjects), reducing statistical power. %
However, our power analysis suggests a minimum of 32 participants per condition for power of 0.8 and a medium effect size, which we exceeded. We also control for exogenous factors using an ANCOVA, highlighting differences between groups. \newt{A final limitation is that these scam conversations may be influenced or constrained by the safety guardrails built into commercial LLMs. While we can work around these restrictions to create conversations we believe are realistic, these models may still be prevented from fully mimicking how real scammers behave, which could reduce the realism of the system.}

There are many ways that this work can be expanded in the future. First, we intend to conduct a longitudinal study on our approach to see which ones lead to greater knowledge retention. As of now, we can only claim the groups teach the information better for a short period of time. Second, an interesting study would be to test if different LLMs result in greater levels of inoculation or if a combination of LLMs would perform best. Third, alternative interaction techniques could be explored such as placing the user in the position of the scammer, with an advisor and target being simulated interlocutors. \newt{Fourth, we would like to test the effectiveness of \textsf{ScamPilot} on different scam types. We would also like to test how the application performs across different demographics to see if this approach works better for groups targeted more frequently by scams.} Finally, future work can explore whether the \textsf{ScamPilot} interaction technique can be used for other HCI or cybersecurity applications.%
\section{Conclusion}
In this paper we presented \textsf{ScamPilot}, a system that simulates scam conversations using inter-agent communication with active user engagement through multiple-choice questions and advice giving. In a between-groups experiment (N = 150), with three experimental conditions and one control, the \textit{quiz+advice} interface increased scam recognition (+8\%), response-efficacy (+9\%), and change in self-efficacy (+19\%) without significantly decreasing legitimate recognition. This result suggests that combining inoculation with active learning is an effective way to increase scam prevention without raising wariness to legitimate information. Through this study, we show the implications of different interfaces in that embedded quizzes cause participants to think they learned more, giving advice increases scam recognition, and the combination of the two is ideal. We discuss the ethics and limitations for the system while highlighting ways it can be used in future settings.

\bibliographystyle{ACM-Reference-Format}
\bibliography{_REFERENCES}

\appendix
\section*{Appendix}
\label{sec:appendix}

\begin{table}[H]
\caption{Demographic information about participants in our study who completed the attention check (N=150).}
\begin{tabular}{llccccc}
\rowcolor[HTML]{FFFFFF} 
{\color[HTML]{333333} } & {\color[HTML]{333333} } & {\color[HTML]{333333} \textbf{Control}} & {\color[HTML]{333333} \textbf{Advice}} & {\color[HTML]{333333} \textbf{Quiz+Advice}} & {\color[HTML]{333333} \textbf{Quiz}} & \cellcolor[HTML]{FFFFFF}{\color[HTML]{333333} \textbf{Overall}} \\ \hline
\rowcolor[HTML]{FFFFFF} 
{\color[HTML]{333333} } & {\color[HTML]{333333} \textbf{Demographic Characteristic}} & {\color[HTML]{333333} \textbf{\#}} & {\color[HTML]{333333} \textbf{\#}} & {\color[HTML]{333333} \textbf{\#}} & {\color[HTML]{333333} \textbf{\#}} & \cellcolor[HTML]{FFFFFF}{\color[HTML]{333333} \textbf{\#}} \\ \hline
\rowcolor[HTML]{EFEFEF} 
\cellcolor[HTML]{FFFFFF}{\color[HTML]{333333} } & {\color[HTML]{333333} \textbf{Female}} & {\color[HTML]{333333} 19} & {\color[HTML]{333333} 15} & {\color[HTML]{333333} 23} & \multicolumn{1}{c|}{\cellcolor[HTML]{EFEFEF}{\color[HTML]{333333} 15}} & \cellcolor[HTML]{EFEFEF}{\color[HTML]{333333} 72} \\
\rowcolor[HTML]{FFFFFF} 
\cellcolor[HTML]{FFFFFF}{\color[HTML]{333333} } & {\color[HTML]{333333} \textbf{Male}} & {\color[HTML]{333333} 14} & {\color[HTML]{333333} 17} & {\color[HTML]{333333} 17} & \multicolumn{1}{c|}{\cellcolor[HTML]{FFFFFF}{\color[HTML]{333333} 26}} & \cellcolor[HTML]{FFFFFF}{\color[HTML]{333333} 74} \\
\rowcolor[HTML]{EFEFEF} 
\cellcolor[HTML]{FFFFFF}{\color[HTML]{333333} } & {\color[HTML]{333333} \textbf{Non-binary}} & {\color[HTML]{333333} 1} & {\color[HTML]{333333} 2} & {\color[HTML]{333333} 0} & \multicolumn{1}{c|}{\cellcolor[HTML]{EFEFEF}{\color[HTML]{333333} 0}} & \cellcolor[HTML]{EFEFEF}{\color[HTML]{333333} 3} \\
\rowcolor[HTML]{FFFFFF} 
\multirow{-4}{*}{\cellcolor[HTML]{FFFFFF}{\color[HTML]{333333} \textbf{Gender}}} & {\color[HTML]{333333} \textbf{Prefer not to say}} & {\color[HTML]{333333} 0} & {\color[HTML]{333333} 0} & {\color[HTML]{333333} 0} & \multicolumn{1}{c|}{\cellcolor[HTML]{FFFFFF}{\color[HTML]{333333} 1}} & \cellcolor[HTML]{FFFFFF}{\color[HTML]{333333} 1} \\ \hline
\rowcolor[HTML]{EFEFEF} 
\cellcolor[HTML]{FFFFFF}{\color[HTML]{333333} } & {\color[HTML]{333333} \textbf{18-24 years}} & {\color[HTML]{333333} 4} & {\color[HTML]{333333} 1} & {\color[HTML]{333333} 2} & \multicolumn{1}{c|}{\cellcolor[HTML]{EFEFEF}{\color[HTML]{333333} 5}} & \cellcolor[HTML]{EFEFEF}{\color[HTML]{333333} 12} \\
\rowcolor[HTML]{FFFFFF} 
\cellcolor[HTML]{FFFFFF}{\color[HTML]{333333} } & {\color[HTML]{333333} \textbf{25-34 years}} & {\color[HTML]{333333} 10} & {\color[HTML]{333333} 11} & {\color[HTML]{333333} 13} & \multicolumn{1}{c|}{\cellcolor[HTML]{FFFFFF}{\color[HTML]{333333} 15}} & \cellcolor[HTML]{FFFFFF}{\color[HTML]{333333} 49} \\
\rowcolor[HTML]{EFEFEF} 
\cellcolor[HTML]{FFFFFF}{\color[HTML]{333333} } & {\color[HTML]{333333} \textbf{35-44 years}} & {\color[HTML]{333333} 8} & {\color[HTML]{333333} 8} & {\color[HTML]{333333} 7} & \multicolumn{1}{c|}{\cellcolor[HTML]{EFEFEF}{\color[HTML]{333333} 11}} & \cellcolor[HTML]{EFEFEF}{\color[HTML]{333333} 34} \\
\rowcolor[HTML]{FFFFFF} 
\cellcolor[HTML]{FFFFFF}{\color[HTML]{333333} } & {\color[HTML]{333333} \textbf{45-54 years}} & {\color[HTML]{333333} 9} & {\color[HTML]{333333} 10} & {\color[HTML]{333333} 9} & \multicolumn{1}{c|}{\cellcolor[HTML]{FFFFFF}{\color[HTML]{333333} 4}} & \cellcolor[HTML]{FFFFFF}{\color[HTML]{333333} 32} \\
\rowcolor[HTML]{EFEFEF} 
\cellcolor[HTML]{FFFFFF}{\color[HTML]{333333} } & {\color[HTML]{333333} \textbf{55-64 years}} & {\color[HTML]{333333} 3} & {\color[HTML]{333333} 2} & {\color[HTML]{333333} 6} & \multicolumn{1}{c|}{\cellcolor[HTML]{EFEFEF}{\color[HTML]{333333} 6}} & \cellcolor[HTML]{EFEFEF}{\color[HTML]{333333} 17} \\
\rowcolor[HTML]{FFFFFF} 
\multirow{-6}{*}{\cellcolor[HTML]{FFFFFF}{\color[HTML]{333333} \textbf{Age}}} & {\color[HTML]{333333} \textbf{65+ years}} & {\color[HTML]{333333} 0} & {\color[HTML]{333333} 2} & {\color[HTML]{333333} 3} & \multicolumn{1}{c|}{\cellcolor[HTML]{FFFFFF}{\color[HTML]{333333} 1}} & \cellcolor[HTML]{FFFFFF}{\color[HTML]{333333} 6} \\ \hline
\rowcolor[HTML]{EFEFEF} 
\cellcolor[HTML]{FFFFFF}{\color[HTML]{333333} } & {\color[HTML]{333333} \textbf{White}} & {\color[HTML]{333333} 24} & {\color[HTML]{333333} 30} & {\color[HTML]{333333} 26} & \multicolumn{1}{c|}{\cellcolor[HTML]{EFEFEF}{\color[HTML]{333333} 32}} & \cellcolor[HTML]{EFEFEF}{\color[HTML]{333333} 112} \\
\rowcolor[HTML]{FFFFFF} 
\cellcolor[HTML]{FFFFFF}{\color[HTML]{333333} } & {\color[HTML]{333333} \textbf{Black or African American}} & {\color[HTML]{333333} 8} & {\color[HTML]{333333} 1} & {\color[HTML]{333333} 11} & \multicolumn{1}{c|}{\cellcolor[HTML]{FFFFFF}{\color[HTML]{333333} 4}} & \cellcolor[HTML]{FFFFFF}{\color[HTML]{333333} 24} \\
\rowcolor[HTML]{EFEFEF} 
\cellcolor[HTML]{FFFFFF}{\color[HTML]{333333} } & {\color[HTML]{333333} \textbf{Asian}} & {\color[HTML]{333333} 2} & {\color[HTML]{333333} 2} & {\color[HTML]{333333} 2} & \multicolumn{1}{c|}{\cellcolor[HTML]{EFEFEF}{\color[HTML]{333333} 4}} & \cellcolor[HTML]{EFEFEF}{\color[HTML]{333333} 10} \\
\rowcolor[HTML]{FFFFFF} 
\cellcolor[HTML]{FFFFFF}{\color[HTML]{333333} } & {\color[HTML]{333333} \textbf{Prefer to self-describe}} & {\color[HTML]{333333} 0} & {\color[HTML]{333333} 1} & {\color[HTML]{333333} 1} & \multicolumn{1}{c|}{\cellcolor[HTML]{FFFFFF}{\color[HTML]{333333} 1}} & \cellcolor[HTML]{FFFFFF}{\color[HTML]{333333} 3} \\
\rowcolor[HTML]{EFEFEF} 
\multirow{-5}{*}{\cellcolor[HTML]{FFFFFF}{\color[HTML]{333333} \textbf{Race/Ethnicity}}} & {\color[HTML]{333333} \textbf{Native Hawaiian/Other Pacific Islander}} & {\color[HTML]{333333} 0} & {\color[HTML]{333333} 0} & {\color[HTML]{333333} 0} & \multicolumn{1}{c|}{\cellcolor[HTML]{EFEFEF}{\color[HTML]{333333} 1}} & \cellcolor[HTML]{EFEFEF}{\color[HTML]{333333} 1} \\ \hline
\rowcolor[HTML]{FFFFFF} 
\cellcolor[HTML]{FFFFFF}{\color[HTML]{333333} } & {\color[HTML]{333333} \textbf{Bachelor's degree}} & {\color[HTML]{333333} 8} & {\color[HTML]{333333} 12} & {\color[HTML]{333333} 12} & \multicolumn{1}{c|}{\cellcolor[HTML]{FFFFFF}{\color[HTML]{333333} 18}} & \cellcolor[HTML]{FFFFFF}{\color[HTML]{333333} 50} \\
\rowcolor[HTML]{EFEFEF} 
\cellcolor[HTML]{FFFFFF}{\color[HTML]{333333} } & {\color[HTML]{333333} \textbf{High school or equivalent}} & {\color[HTML]{333333} 12} & {\color[HTML]{333333} 7} & {\color[HTML]{333333} 14} & \multicolumn{1}{c|}{\cellcolor[HTML]{EFEFEF}{\color[HTML]{333333} 11}} & \cellcolor[HTML]{EFEFEF}{\color[HTML]{333333} 44} \\
\rowcolor[HTML]{FFFFFF} 
\cellcolor[HTML]{FFFFFF}{\color[HTML]{333333} } & {\color[HTML]{333333} \textbf{Master's degree}} & {\color[HTML]{333333} 5} & {\color[HTML]{333333} 10} & {\color[HTML]{333333} 9} & \multicolumn{1}{c|}{\cellcolor[HTML]{FFFFFF}{\color[HTML]{333333} 7}} & \cellcolor[HTML]{FFFFFF}{\color[HTML]{333333} 31} \\
\rowcolor[HTML]{EFEFEF} 
\cellcolor[HTML]{FFFFFF}{\color[HTML]{333333} } & {\color[HTML]{333333} \textbf{Associate degree}} & {\color[HTML]{333333} 7} & {\color[HTML]{333333} 5} & {\color[HTML]{333333} 2} & \multicolumn{1}{c|}{\cellcolor[HTML]{EFEFEF}{\color[HTML]{333333} 4}} & \cellcolor[HTML]{EFEFEF}{\color[HTML]{333333} 18} \\
\rowcolor[HTML]{FFFFFF} 
\cellcolor[HTML]{FFFFFF}{\color[HTML]{333333} } & {\color[HTML]{333333} \textbf{Prefer to self-describe}} & {\color[HTML]{333333} 0} & {\color[HTML]{333333} 0} & {\color[HTML]{333333} 2} & \multicolumn{1}{c|}{\cellcolor[HTML]{FFFFFF}{\color[HTML]{333333} 0}} & \cellcolor[HTML]{FFFFFF}{\color[HTML]{333333} 2} \\
\rowcolor[HTML]{EFEFEF} 
\cellcolor[HTML]{FFFFFF}{\color[HTML]{333333} } & {\color[HTML]{333333} \textbf{Doctorate degree}} & {\color[HTML]{333333} 1} & {\color[HTML]{333333} 0} & {\color[HTML]{333333} 1} & \multicolumn{1}{c|}{\cellcolor[HTML]{EFEFEF}{\color[HTML]{333333} 0}} & \cellcolor[HTML]{EFEFEF}{\color[HTML]{333333} 2} \\
\rowcolor[HTML]{FFFFFF} 
\cellcolor[HTML]{FFFFFF}{\color[HTML]{333333} } & {\color[HTML]{333333} \textbf{Professional degree}} & {\color[HTML]{333333} 1} & {\color[HTML]{333333} 0} & {\color[HTML]{333333} 0} & \multicolumn{1}{c|}{\cellcolor[HTML]{FFFFFF}{\color[HTML]{333333} 1}} & \cellcolor[HTML]{FFFFFF}{\color[HTML]{333333} 2} \\
\rowcolor[HTML]{EFEFEF} 
\multirow{-8}{*}{\cellcolor[HTML]{FFFFFF}{\color[HTML]{333333} \textbf{Education Level}}} & {\color[HTML]{333333} \textbf{Some formal education}} & {\color[HTML]{333333} 0} & {\color[HTML]{333333} 0} & {\color[HTML]{333333} 0} & \multicolumn{1}{c|}{\cellcolor[HTML]{EFEFEF}{\color[HTML]{333333} 1}} & \cellcolor[HTML]{EFEFEF}{\color[HTML]{333333} 1} \\ \hline
\rowcolor[HTML]{FFFFFF} 
{\color[HTML]{333333} \textbf{Total}} & {\color[HTML]{333333} \textbf{}} & {\color[HTML]{333333} 34} & {\color[HTML]{333333} 34} & {\color[HTML]{333333} 40} & \multicolumn{1}{c|}{\cellcolor[HTML]{FFFFFF}{\color[HTML]{333333} 42}} & \cellcolor[HTML]{FFFFFF}{\color[HTML]{333333} 150} \\ \hline
\end{tabular}
\label{table:demographics}
\end{table}

\begin{table}[t]
\centering
\caption{Advice Codebook (Themes, Categories, Codes, Definitions, and Examples)}
\label{tab:advicecodebook}

\begingroup
\setlength{\tabcolsep}{4.5pt}      
\renewcommand{\arraystretch}{1.12} 

\begin{adjustbox}{max totalsize={\textwidth}{0.92\textheight},center} 
\begin{tabularx}{1.15\textwidth}{%
  >{\bfseries}p{0.16\textwidth}
  >{\bfseries}p{0.18\textwidth}
  >{\bfseries}p{0.22\textwidth}
  >{\RaggedRight\arraybackslash}X
  >{\RaggedRight\arraybackslash}X}

\toprule
Theme & Category & Code & Type of Advice & Examples \\
\midrule

\rowcolor{ThemeGreen}
\multicolumn{5}{>{\RaggedRight\arraybackslash}p{\dimexpr\linewidth-2\tabcolsep\relax}}{\textbf{Type 1: Advice Encouraging \newt{Verification}}}\\
\addlinespace[3pt]

& \multirow{3}{=}{Direct identity verification}
  & Ask for specific details
  & Request information only the real person would know (names, events, inside knowledge).
  & Ask another question that only Mark would know. \\
&   & Request real-time photo or video
    & Ask for a live photo or video to confirm identity or context.
    & Tell him to take a selfie of himself and the background. \\
&   & Direct call
    & Start a phone or video call to verify sender's identity and legitimacy.
    & If grandson has a phone, call it and see if he answers. If he does then this is a scam. \\

& \multirow{2}{=}{Checking with trusted sources}
  & Contact family member or friend
  & Reach out to friend or relative (parent, sibling, \ldots) to verify or get help.
  & Insist on putting a call to his parents. \\
&   & Ask/involve third party
    & Advise to involve trustworthy external parties (police, staff, company reps) to verify or get support.
    & If you don't want to involve your parents please call 911 and have the police help out. \\

& \multirow{2}{=}{Verify situation}
  & Request more information
  & Ask for clarifying details (location, people involved, reasons for request).
  & Ask him what the subway fare is now. \\
&   & Provide alternative safe option
    & Suggest safer alternative action or meeting plan that avoids risk or direct engagement with the suspicious request.
    & Tell me the airport you will be using and I will buy the ticket for you. \\
\addlinespace[3pt]

\rowcolor{ThemePink}
\multicolumn{5}{>{\RaggedRight\arraybackslash}p{\dimexpr\linewidth-2\tabcolsep\relax}}{\textbf{Type 2: Advice \newt{Encouraging Refusals}}}\\
\addlinespace[3pt]

& \multirow{4}{=}{Strategy of information sharing and communication}
  & Limit personal information
  & Discourage sharing sensitive or private details (personal data, address).
  & Don't provide any personal information. This could be a scam. \\
&   & Refuse of money transfer
    & Discourage sending money.
    & Never send money to anyone through a text or email request. \\
&   & Refuse to take action
    & Stop communication or do not comply with any instructions.
    & Stop talking to him Jane. It is pointless. \\
&   & Refuse under condition
    & Refuse action until verification occurs.
    & Grandma, please don't engage with him any further until you have called Mark's number and verified that he hasn't answered his phone. \\
\addlinespace[3pt]

\rowcolor{ThemePurple}
\multicolumn{5}{>{\RaggedRight\arraybackslash}p{\dimexpr\linewidth-2\tabcolsep\relax}}{\textbf{Type 3: \newt{Advice on Managing the Situation}}}\\
\addlinespace[3pt]

& \multirow{3}{=}{Emotional \& psychological feedback}
  & Recognize manipulation
  & Identify emotional tactics such as urgency, fear, or emotional appeals.
  & Jane, the scammer is begging and acting a fool. They are applying pressure and trying to scam you at all costs. It is manipulation and a widely known tactic to play with your emotions. \\
&   & Offer emotional support
    & Offer reassurance to calm down and encourage rational thinking.
    & Good job asking a question only Mark would know! \\
&   & Urge caution
    & Increase awareness and skepticism.
    & Be careful, the scammers will impersonate relatives in a crisis. Wait, do not be in a hurry—find out whether it actually IS Mark. \\
\addlinespace[3pt]

\rowcolor{ThemeBlue}
\multicolumn{5}{>{\RaggedRight\arraybackslash}p{\dimexpr\linewidth-2\tabcolsep\relax}}{\textbf{Type 4: Advice \newt{on Pointing Out Flaws}}}\\
\addlinespace[3pt]

& \multirow{3}{=}{Skepticism \& logical reasoning}
  & Identify contradictions
  & Highlight inconsistencies in the story or behavior.
  & Mark just said \$200 and then \$300, this is suspicious. Also, tickets in Boston don't cost this much. \\
&   & Identify being generic
    & Point out vague wording or language that is not personalized.
    & The message is lacking in specifics. \\
&   & Provide alternative fact
    & Offer a plausible alternative explanation to challenge what the sender claims (``This could be found on social media'').
    & Let them know you never watched that movie with them. \\
\addlinespace[3pt]

\textbf{Not Relevant}
& —
& —
& The advice lacks relevance on how to proceed or verify the situation.
& I cannot help you. \\

\bottomrule
\end{tabularx}
\end{adjustbox}
\endgroup
\end{table}

\begin{table}[t]
\centering
\caption{The six attention check questions that participants answered throughout their session.}
\label{tab:attentioncheck}

\begingroup
\setlength{\tabcolsep}{4.5pt}      
\renewcommand{\arraystretch}{1.12}

\begin{adjustbox}{max totalsize={\textwidth}{0.92\textheight},center}
\begin{tabularx}{1.05\textwidth}{
  >{\bfseries}p{0.20\textwidth}  
  >{\bfseries}p{0.25\textwidth}  
  >{\RaggedRight\arraybackslash}X}
\toprule
Number & Question & Options \\
\midrule

\rowcolor{ThemeGreen}
\multicolumn{3}{>{\RaggedRight\arraybackslash}p{\dimexpr\linewidth-2\tabcolsep\relax}}%
{\textbf{Pre-Survey Attention Check Questions}} \\
\addlinespace[3pt]

1. MCQ & If you are reading this, please select agree. & [Strongly disagree, Disagree, Neutral, Agree, Strongly agree] \\

2. Likert Scale & If you are reading this, choose Disagree. & [Strongly disagree, Disagree, Neither agree nor disagree, Agree, Strongly agree]\\
\addlinespace[3pt]

\rowcolor{ThemePink}
\multicolumn{3}{>{\RaggedRight\arraybackslash}p{\dimexpr\linewidth-2\tabcolsep\relax}}%
{\textbf{Tutorial Video Attention Check Question}} \\
\addlinespace[3pt]

3. MCQ & What type of scam was discussed in the video? & [Job Scams, Pig Butchering Scams, Online Shopping Scams, Imposter Scams] \\

\addlinespace[3pt]

\rowcolor{ThemeBlue}
\multicolumn{3}{>{\RaggedRight\arraybackslash}p{\dimexpr\linewidth-2\tabcolsep\relax}}%
{\textbf{Post-Survey Attention Check Questions}} \\
\addlinespace[3pt]

4. Likert Scale & If you are reading this, select Disagree & [Strongly disagree, Disagree, Neither agree nor disagree, Agree, Strongly agree] \\
5. Likert Scale & You are contacted by somebody saying they work for the IRS. They want to know if you are reading this survey and if so select Either for this question. How would you respond? & [Extremely Unlikely, Unlikely, Slightly Unlikely, Either, Slightly Likely, Likely,
Extremely Likely] \\
6. Likert Scale + Justification & 
How likely is it that this message is a scam?
\begin{minipage}[c]{\linewidth}
    \centering
    \includegraphics[width=0.8\linewidth]{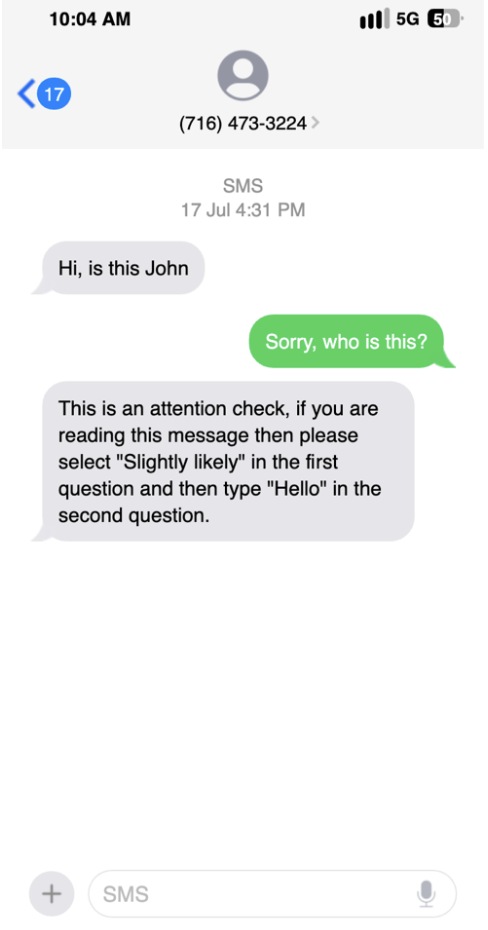} 
\end{minipage} Briefly justify your response to the previous question.
& 
Likert scale: [Extremely unlikely, Unlikely, Slightly Unlikely, Either, Slightly Likely, Likely, Extremely Likely]
\\
\bottomrule
\end{tabularx}
\end{adjustbox}
\endgroup
\end{table}

\end{document}